\documentclass[twocolumn]{aastex631}
\usepackage{graphicx}

\received{December 19, 2024}
\accepted{January 10, 2025}

\shorttitle{The influence of chromospheric activity}
\shortauthors{Vieytes et al.}

\graphicspath{{./}{figures/}}

\begin{document}

\title{The influence of chromospheric activity on line formation}

\correspondingauthor{Mariela C. Vieytes}
\email{mvieytes@untref.edu.ar}

\author[0000-0003-4615-8746]{Mariela C. Vieytes}
\affiliation{Instituto de Astronomía y Física del Espacio,(IAFE, CONICET-UBA) 
Buenos Aires, Argentina }
\affiliation{Departamento de Ciencia y Tecnología, UNTREF, Buenos Aires, Argentina}
\affil{Center for Computational Astrophysics (CCA), Flatiron Institute, New York, NY, USA} 

\author[0000-0002-3852-3590]{Lily L. Zhao}
\thanks{NASA Sagan Fellow} 
\affil{Department of Astronomy \& Astrophysics, University of Chicago, Chicago, IL, USA}
\affil{Center for Computational Astrophysics (CCA), Flatiron Institute, New York, NY, USA} 

\author[0000-0001-9907-7742]{Megan Bedell}
\affil{Center for Computational Astrophysics (CCA), Flatiron Institute, New York, NY, USA}



\begin{abstract}
One of the primary sources of stellar spectral variability is magnetic activity. While our current understanding of chromospheric activity is largely derived from specific lines sensitive to chromospheric heating, such as the Ca II H\&K doublet, previous observational studies have shown that other spectral lines are also affected. To investigate the influence of activity on line formation in greater detail, we constructed a set of stellar models for hypothetical G2 dwarf stars with varying levels of activity and calculated their synthetic spectra. A comparison of these spectra revealed two spectral regions most significantly impacted by activity: approximately 3300–4400 \AA~ and 5250–5500 \AA. By calculating the total contribution function of the lines, we determined that the emergence of a secondary chromospheric contribution to line formation is the primary mechanism driving these changes. Based on our calculations and analysis, we compiled a list of transition lines and their corresponding changes due to chromospheric activity. This list could serve as a valuable tool for selecting spectral lines applicable to a wide range of astrophysical studies.

\end{abstract}



\section{Introduction} \label{sec:intro}

A comprehensive understanding of stellar spectra is essential for advancing a variety of fields where spectroscopic analysis plays a critical role, such as stellar abundance determination or exoplanet detection and characterization. Magnetic activity is one of the primary drivers of stellar variability, making its characterization crucial.

It is well known that the Ca II H\&K doublet (3968.469 \AA ~and 3933.663 \AA ~respectively) in the visible violet spectral region is sensitive to solar chromospheric activity. On the solar disk, mechanical heating in chromospheric plage regions is the main factor that enhances the emission in the cores of these lines compared to quiet regions. Recently, \cite{crea24} demonstrated that 70 \% of the Ca II variation can be attributed to plages.

Analyzing full-disk observations of the Sun, \citet{liv07} showed that the Ca II K line tracks the solar cycle according to sunspot number, with a peak-to-peak variation in amplitude of about 25\%. They found that the K3 feature (the center of the Ca II K line profile) also experiences variation with the solar cycle. Regarding other lines, they reported a certain degree of change in the core of strong Fe I lines, although the authors suggested further investigation.

\begin{figure*}[t]
\centering
\resizebox{\hsize}{!}{\includegraphics[trim=1cm 1cm 1cm 1cm, clip, width=0.8\textwidth]{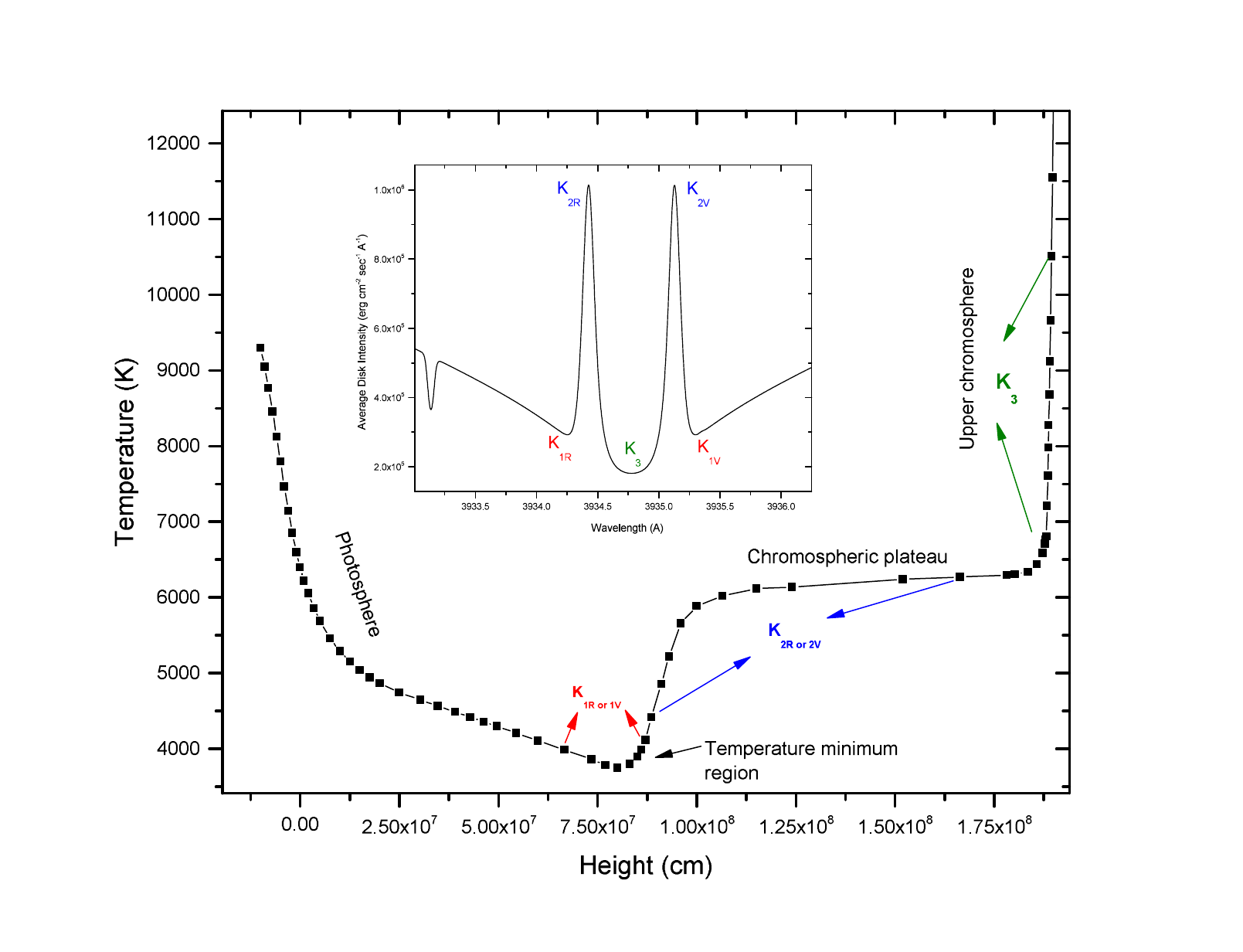}}
\caption{A model for the solar magnetic inter-network structure, the main model component describing the quiet Sun atmosphere, and its calculated Ca II K profile. The arrows indicate the formation region for the K1, K2 (red and blue) and K3 features.}
\label{fig:model_ca2}
\end{figure*}

On the stellar side, the pioneering HK Project at Mount Wilson Observatory made use of the variability in the Ca II H\&K observations to study the stellar chromospheric activity \citep{bal95,wil78}. They established a chromospheric proxy using these line profiles, the $S_{CaII}$ index (and its derivative index R’$_{HK}$), which is commonly used to quantify magnetic variability and the existence of stellar cycles in late-type stars \citep[see for example][]{met13,flor16,bus20}.

Thus, much of our current observational knowledge about chromospheric activity is derived from the Ca II H\&K lines. However, these features are not the only place that chromospheric activity affects the spectrum. More recent work, largely motivated by the importance of stellar spectral variability as a noise source for exoplanet searches, has sought to characterize stellar activity in other spectral features. 

 \citet{wise18} developed a method to find new activity indicators. From a set of spectra obtained by the High Accuracy Radial velocity Planet Searcher (HARPS) project of $\epsilon$ Eridani and $\alpha$ Cen B, the authors generated a list containing 43 activity-sensitive lines to be used by planet hunters to disentangle the information of spots, plages, and activity cycles contained in the stellar spectrum.

\citet{spi20} analyzed high-resolution spectra of 211 sun-like stars of different levels of chromospheric activity and found that the equivalent widths of lines of different neutral and first ionized atoms can increase as a function of the activity index R’$_{HK}$ during the stellar cycle. These outcomes have implications in several astrophysical studies, including radial velocity (RV) measurements for planet hunting.  

From these works and others, it is clear that magnetic activity affects many parts of the stellar spectrum in complex ways. Properly understanding and modeling these effects requires knowledge of the stellar atmosphere and the physics that shape the spectrum.

In the context of using RV measurements to search for exoplanets, \citet{almou22} were able to incorporate a greater degree of stellar physics by investigating the relationship between the formation temperature of a set of line profiles and the resultant measured RV. They showed that the formation temperature is a more consistent diagnostic tool than the straightforward line depth to characterize the impact of photospheric stellar activity on measured RVs at both rotational and magnetic cycle timescales. However, their models do not consider the presence of a chromosphere where the core of strong lines are formed. A more complex calculation accounting for departures from local thermodynamic equilibrium (LTE) would be required to properly obtain the population of the species whose transition lines are formed in the stellar chromosphere. 

Models of the solar surface convection with time-dependent 3D hydrodynamics were used by \citet{dra23}. They generated synthetic LTE spectra for the Sun observed as a star to study the impact of granulation on RV measurements. Nevertheless, most of the models presented in this work were limited to the nonmagnetic photosphere; they did not consider magnetic features responsible for variability over longer timescales, such as activity cycles. 

More recently, \cite{dra24} investigated systematic line-strength variability with chromospheric activity of lines of Fe I, Fe II, Mg I, Mn I, H$\alpha$, H$\beta$, H$\gamma$, Na I, and the G-band. They used three years of HARPS in the North (HARPS-N) observations of the Sun-as-a-star. They observed that Fe II in the blue  exhibited variability with greater amplitude, while the trends changed sign among line in the green Mg I triplet and Balmer lines. Additionally, they found that variation in the G-band core was greater than in the full G-band, as predicted. Beyond their results, the first three sections of their work offer an excellent review on the subject for the reader.

\citet{viey05,viey09} built models of the chromosphere of a selected sample of G and K dwarf stars to study the differences in the thermal structure of their chromospheres with stellar activity.
The models were constructed using the Pandora code \citep{avre03} to obtain the best possible match with the Ca II K and the H$\beta$ observed profiles, considering that these lines are sensitive to magnetic activity. The population of several important atomic species were calculated in non-local thermodynamic equilibrium (NLTE). The results obtained showed that for the less active G stars, the changes with activity were in the region of the temperature minimum, while the most active stars showed changes all along their atmospheric structures, mainly in the upper chromosphere. For the K dwarfs, the changes were produced all along their chromosphere, from the region of the temperature minimum to the transition region, and mainly in the chromospheric plateau, independent of the activity level of the star. The integrated chromospheric radiative losses, normalized to the surface luminosity, showed a unique trend for G and K dwarfs when plotted against $S_{CaII}$. These findings may indicate that the same physical processes are heating the stellar chromospheres in both cases. These works strengthen the capacity of using activity-sensitive chromospheric lines to model the thermal structure of this atmospheric region, and opens a pathway to studying the formation of other lines with activity in solar-type stars.

Stellar signals are now the largest source of error for RV measurements.  Precision RV surveys with the goal of detecting Earth-like planets orbiting around G or K dwarf stars are beginning to reach time baselines on the order of the magnetic cycle timescale; line variations are likely to become even more apparent over these long time baselines. It is therefore becoming essential to distinguish the effect of the changing chromosphere over the visible line profiles in detail, in particular taking into account that the characteristic formation temperature of a line could be modified by the chromospheric activity, and even by its existence.

In the present work, we built a set of atmospheric models with increasing temperature in their chromosphere to simulate the mechanical heating produced by the magnetic activity in dG2 stars. With these synthetic visible spectra, we studied the changes in line formation with activity to disentangle the possible impact of these changes to RV measurements and stellar parameter determinations.

This paper is organized as follows: Section \ref{sec:models} describes the set of stellar atmospheric models constructed for this work. Section \ref{sec:atomic} details the improved Fe I atomic model used to calculate the synthetic spectra. Section \ref{sec:diff} presents the differences observed in the synthetic spectra obtained from the models. Here, we interpret the results using the total contribution function of the computed lines, and relate our findings to previous observational work. Finally, in Section \ref{sec:conclu} we summarizes our outcomes and conclusions.

\begin{figure*}[t]
\centering
\includegraphics[trim=1.5cm 1cm 3cm 1cm, clip, width=0.5\textwidth]{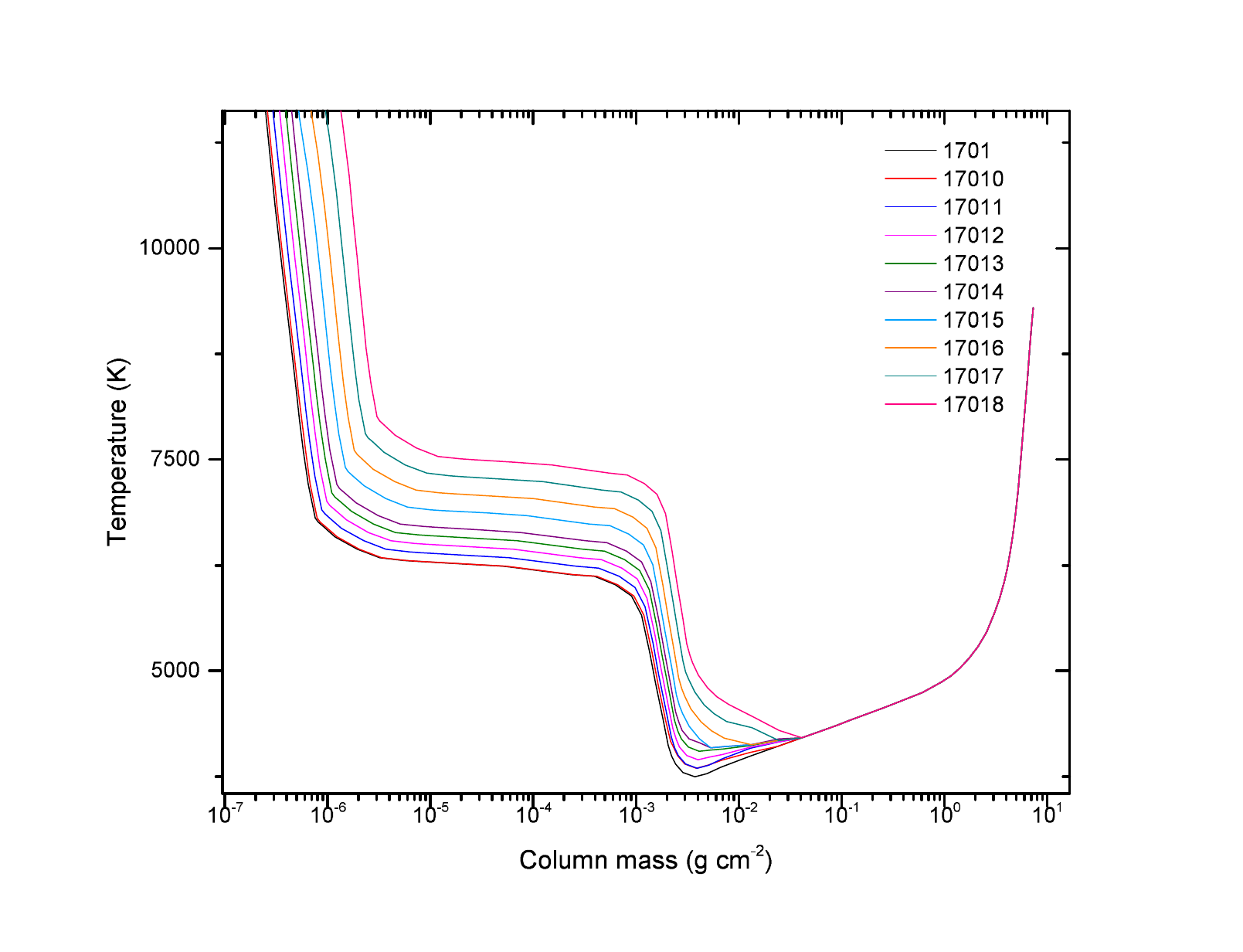}
\includegraphics[trim=1.5cm 1cm 3cm 1cm, clip, width=0.5\textwidth]{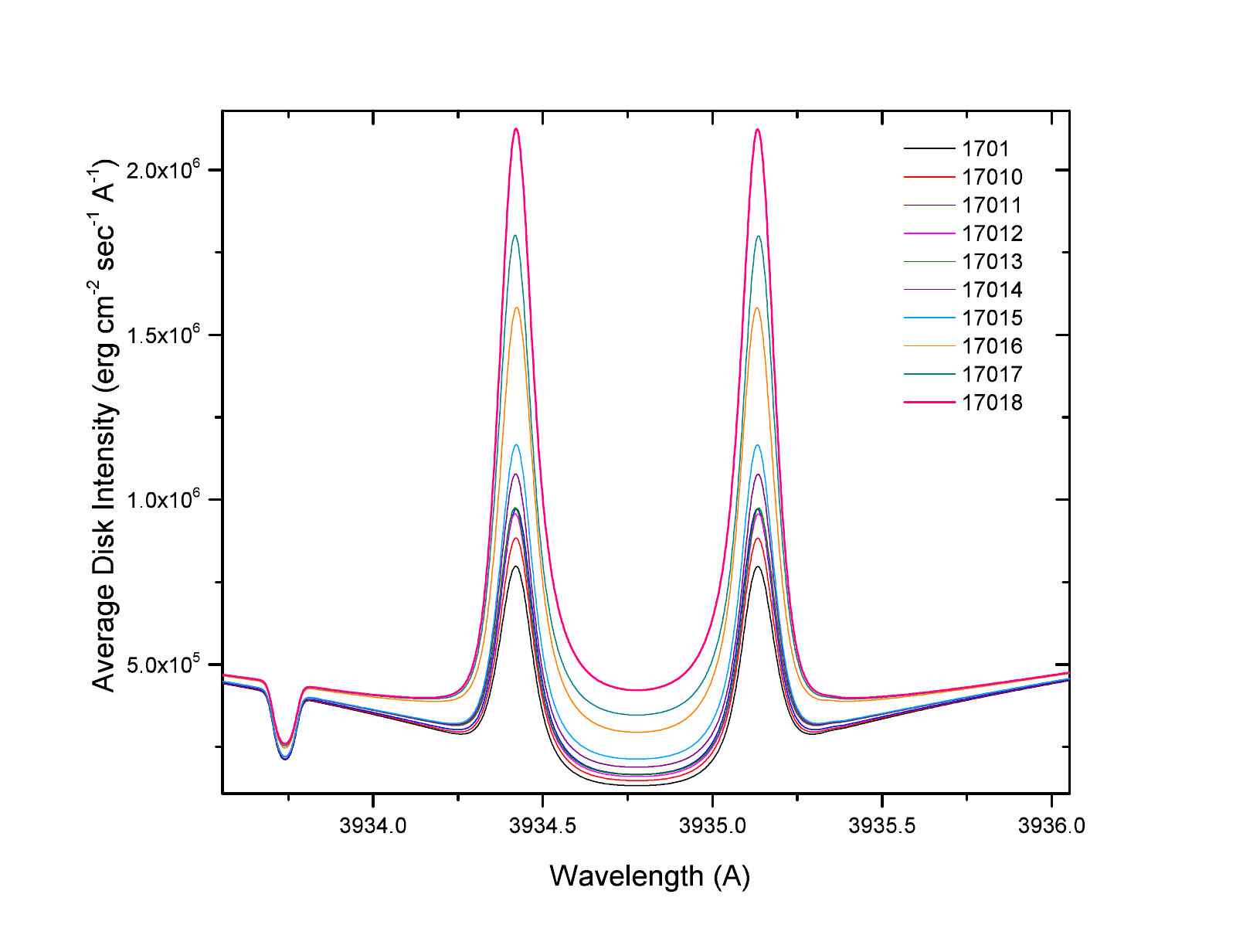}
\caption{\textbf{Left:} Set of atmospheric models with different chromospheric level of activity. \textbf{Right:} Ca II K profiles (vacuum wavelength) calculated from each model, plotted with the same line color that the corresponding model.  }
\label{fig:modelsyca2}
\end{figure*}

\section{The Set of atmospheric models} \label{sec:models}

Following the results obtained by \citet{viey05,viey09}, we know that the thermal structure of the chromosphere of late-type stars can be inferred from the observed Ca II K line profile. This line is excited by collision with electrons, which is a process that is temperature-dependent. In 1D NLTE models, each part of the line forms at different heights in the atmosphere, making it possible to associate a temperature value to atmospheric height. Figure \ref{fig:model_ca2} shows the synthetic, disk-integrated line profile of the Ca II K line (inset) and the approximate corresponding formation region from the model that generated the profile. The model shown in the figure is the 1401 solar model constructed by \citet{fon15} for the solar magnetic inter-network structure. This model serves as the primary component for describing the quiet Sun atmosphere. It is crucial to note that the 1401 model was specifically designed to reproduce a wide range of solar observations and achieves a remarkable agreement with the Solar Spectral Irradiance, as demonstrated in Figure 8 of that work.

All the models calculated in this work were built using the Solar Stellar Radiation Physical Modeling (SSRPM) code library \citep[][and references
therein]{fon16}. The SSRPM assumes hydrostatic equilibrium and solves the statistical equilibrium and radiative transfer equations self-consistently for an atmosphere with plane-parallel or spherical symmetry. 
For the NLTE atomic population calculation, the thermal structure for the photosphere and chromosphere is computed as optically thick, including partial redistribution (PRD). 
The code contains 52 neutral and low ionization state atomic species (generally up to Z$^2$$^+$), H, 
H$^-$, and H$_{2}$. 
Additionally, it allows the calculation of 198 highly ionized species using the optically thin atmosphere approximation. The SSRPM also calculates molecule formation in LTE, which includes molecular sequestration of elements. For spectral lines, 435\,986 transitions are included, along with more than 2\,000\,000 molecular lines of the 20 most abundant and important diatomic molecules for dM stars (e.g., TiO). 

Starting from the quiet solar model 1401, which we call model 1701 here, we built a set of models by increasing the temperature of their chromospheres, but maintaining the same photospheric structure. This method has previously been used in \citet{viey04} and \citet{andre95} to simulate stars with different levels of chromospheric activity. In these 
previous works, the generation of more active chromospheres is done by shifting the same thermal structure inward, i.e.\ towards higher column mass densities. In this case, we increase the temperature with height, which is equivalent to increasing the amount of emitting material at a given chromospheric temperature.
This increase in temperature represents the mechanical energy deposition that strengthens the emission of the Ca II K line core. In other words, increasing temperature can be thought of as increasing the level of simulated magnetic activity for the modeled star; the highest temperature also corresponds to the most active model. Figure \ref{fig:modelsyca2} (Left) shows the thermal structure of the set of models built in this work. The Ca II K profiles resulting from the set of models are shown in the right subplot. The models are named by adding a number at the end of the initial model 1701, with the numbers increasing with the chromospheric activity of the model. 

For each model, we calculated the NLTE population of the species present in the atmosphere \citep[see Table 2 of][]{fon15}, and the resulting visible spectrum from 3300 to 7020 \AA\ (vacuum wavelength). 

\begin{deluxetable*}{ccccc}
\tablecaption{Characteristics of the set of models. First column: name of the model. Second column: The model's minimum temperature. Third column: Atmospheric height of the temperature minimum. Fourth column: Temperature of the plateau for a height of 1.24x10$^{8}$ cm. Fifth column: Percentage Relative Difference of the Integrated Ca II K line flux for each model ($IF_{model}$) with that of model 1701. \label{tab:models}}
\tablewidth{0pt}
\tablehead{
\colhead{ Model name} & \colhead{T$_{min}$}  & \colhead{Height $T_{min}$ } & \colhead{T$_{plateau}$} & \colhead{RD for $IF_{model}$ }\\
 \colhead{} & \colhead{ (K)}  & \colhead{(10$^{7}$cm)} &  \colhead{ (K)}  & \colhead{ (\%)
 }}
\decimalcolnumbers
\startdata
1701  & 3750 & 8.00 & 6140 & 0 \\
17010 & 3850 & 8.00 & 6140 & 10.46 \\
17011 & 3850 & 8.00 & 6240 & 18.98\\
17012 & 3950 & 8.00 & 6340 & 19.01\\
17013 & 4050 & 8.00 & 6440 & 21.19\\
17014 & 4095 & 7.70 & 6540 & 29.01\\
17015 & 4095 & 7.70 & 6740 & 44.46\\
17016 & 4130 & 6.66 & 6940 & 92.22\\
17017 & 4180 & 6.00 & 7140 & 118.64\\
17018 & 4210 & 5.45 & 7340 & 155.74 \\
\enddata
\end{deluxetable*}

Table \ref{tab:models} presents the models' characteristics. The first column shows the model name. To characterize the level of activity of each model, we determine the temperature at two select heights (or column-mass) in the atmosphere easily identified in Figure \ref{fig:modelsyca2}: the temperature of the minimum (second column), and the temperature at the chromospheric plateau (fourth column) corresponding to a height of 1.24x10$^{8}$ cm (which is equivalent to a column-mass from 2.37x10$^{-4}$ g/cm$^{-2}$, and 5.49x10$^{-4}$ in models 1701 and 17018 respectively).The third column gives the location of the temperature minimum. Although the temperature at this height could be the same in some models, its location may shift inward while preserving the shape of the thermal structure of this region, consistent with the results found by \cite{viey09}.  

Model 17010 was constructed such that it differs from model 1701 only in the minimum temperature. Starting from model 17011, increases were made in either the temperature or the location of the temperature minimum region. Additionally, a fixed increase in the plateau temperature was applied from model 17011 onward. 

The last column of Table~\ref{tab:models} provides the Relative Difference (RD) in percentage of the integrated flux of the Ca II K line profile for each model ($IF_{model}$) compared to model 1701 ($IF_{1701}$), as 
\begin{equation}
\frac{\left( IF_{\text{model}} - IF_{1701} \right) \cdot 100}{IF_{1701}}
\label{ec:rd_ca}
\end{equation}

The integration was performed between the $K_1$ minima on both sides of the line as a measure of the activity level (See Figure \ref{fig:modelsyca2}, Right). It is important to note that the Ca II K profile slightly broadens as the emission increases, causing the position of the $K_1$ minima between which the integration is performed to vary slightly for each model. Nevertheless, the distance between these $K_1$ minima remains approximately 1 \AA ~width for all models. This integration is analogous to the commonly used Ca II K solar activity index \citep{don94}. Therefore, the last column of Table~\ref{tab:models} represents the increase in the Ca II K index with activity. 

\cite{ege17} demonstrated a linear relationship between the $S_{CaII}$ and the Ca II K indices for the Sun. Based on this findings, the RD values reported in Table~\ref{tab:models} can be interpreted as the RD in the $S_{CaII}$ index between models. Using their Equation 12 for Solar Cycle 23, the RD between the Ca II K index (or $S_{CaII}$ index) at solar maximum and minimum is approximately 9.2\%, which is a slightly lower than the difference observed between our first two models. \cite{red24} presented monthly averages of the Ca II K index for several Solar Cycles (their Figure 1), showing RD values between solar maximum and minimum ranging from approximately 9.7\% (Solar Cycle 24) to 19.5\% (Solar Cycle 22), values encompassed by models 1701 to 17013 in Table~\ref{tab:models}.

\begin{figure}[ht]
\centering
\resizebox{\hsize}{!}{\includegraphics[trim=2cm 1cm 1cm 2cm, clip, width=1.0\textwidth]{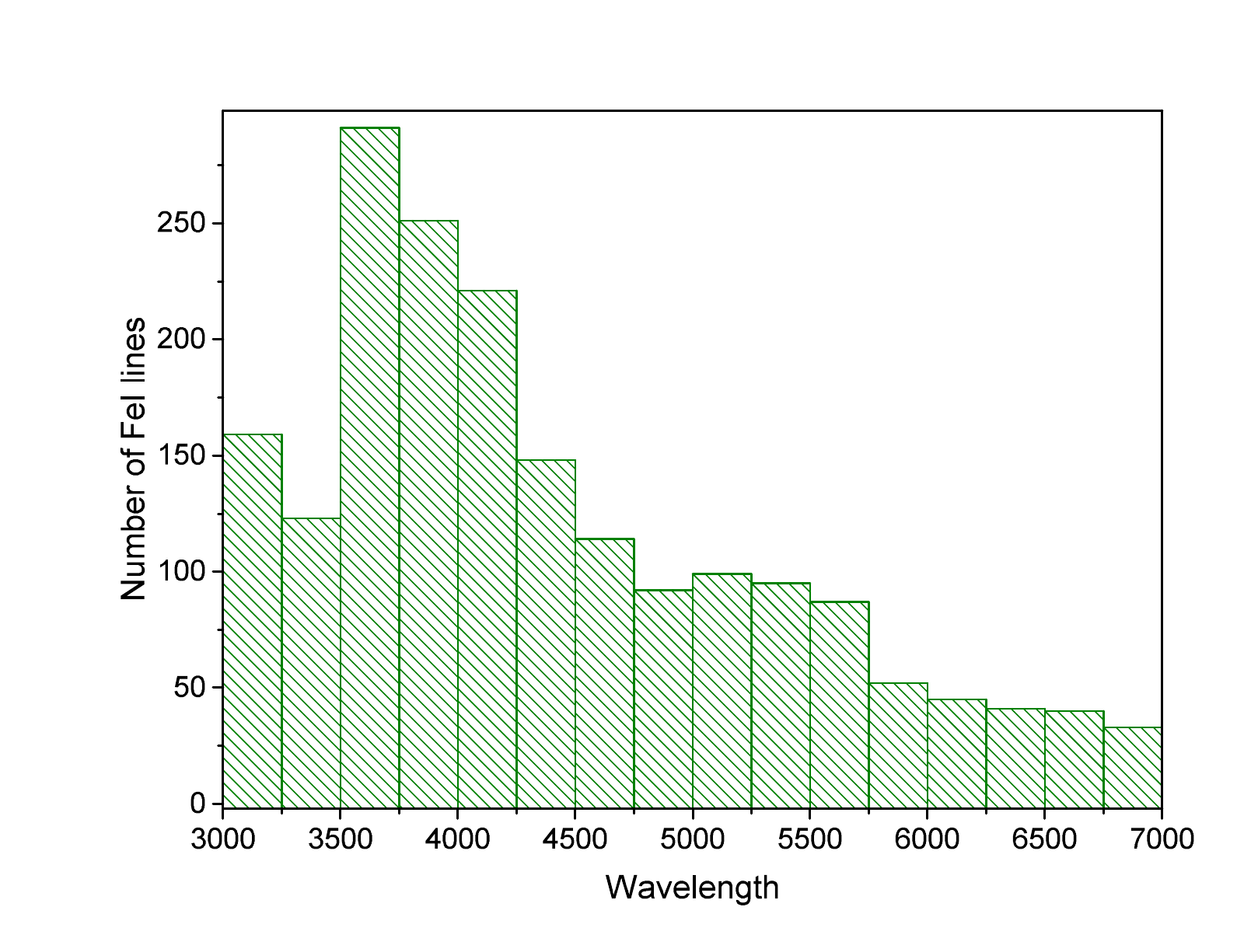}}
\caption{Distribution of the Fe I lines calculated in NLTE with the new atomic model between 3000 and 7000 \AA.}
\label{fig:histogram}
\end{figure}

\section{The improved Fe I atomic model} \label{sec:atomic}
The atomic models used to calculate the population in NLTE are the same as those described in \cite{fon15} and references therein, with the exception of the Fe I atomic model.

It is well known that the most abundant lines in the visible spectra of late-type stars are Fe I lines. For this reason, we decided to update the atomic model for this species. The previous atomic model was assembled by the first 120 energy levels, and their 381 corresponding sublevels, until reaching an energy of 54386.189 cm$^{-1}$. The data were taken from the National Institute of Standards and Technology (NIST) database version 2004 \citep{nist04}. The original \cite{fon15} solar model, which they named model 1401, was calculated with this atomic model that included 1625 Fe I lines, 1368  within the 3300–7020 \AA.

We expanded and updated the atomic structure using the latest version of the NIST database  \citep{nist22}.The new Fe I model was constructed to include levels and sublevels up to a similar energy as the previous atomic model (54379.384 cm$^{-1}$), but with an increased number of levels, now totaling 141 energy levels, and 444 corresponding sublevels.
From the same NIST version, we selected 2397 Fe~I lines with well-determined oscillator strengths, 1715 of which fall within the 3300-7020 \AA~range. The distribution of these lines within the spectral range of interest is shown in Figure \ref{fig:histogram}. For more details about this new atomic model, see \citet{viey24}.

\begin{figure}[t]
\centering
\resizebox{\hsize}{!}{\includegraphics[trim=2cm 1cm 1cm 2cm, clip, width=1.0\textwidth]{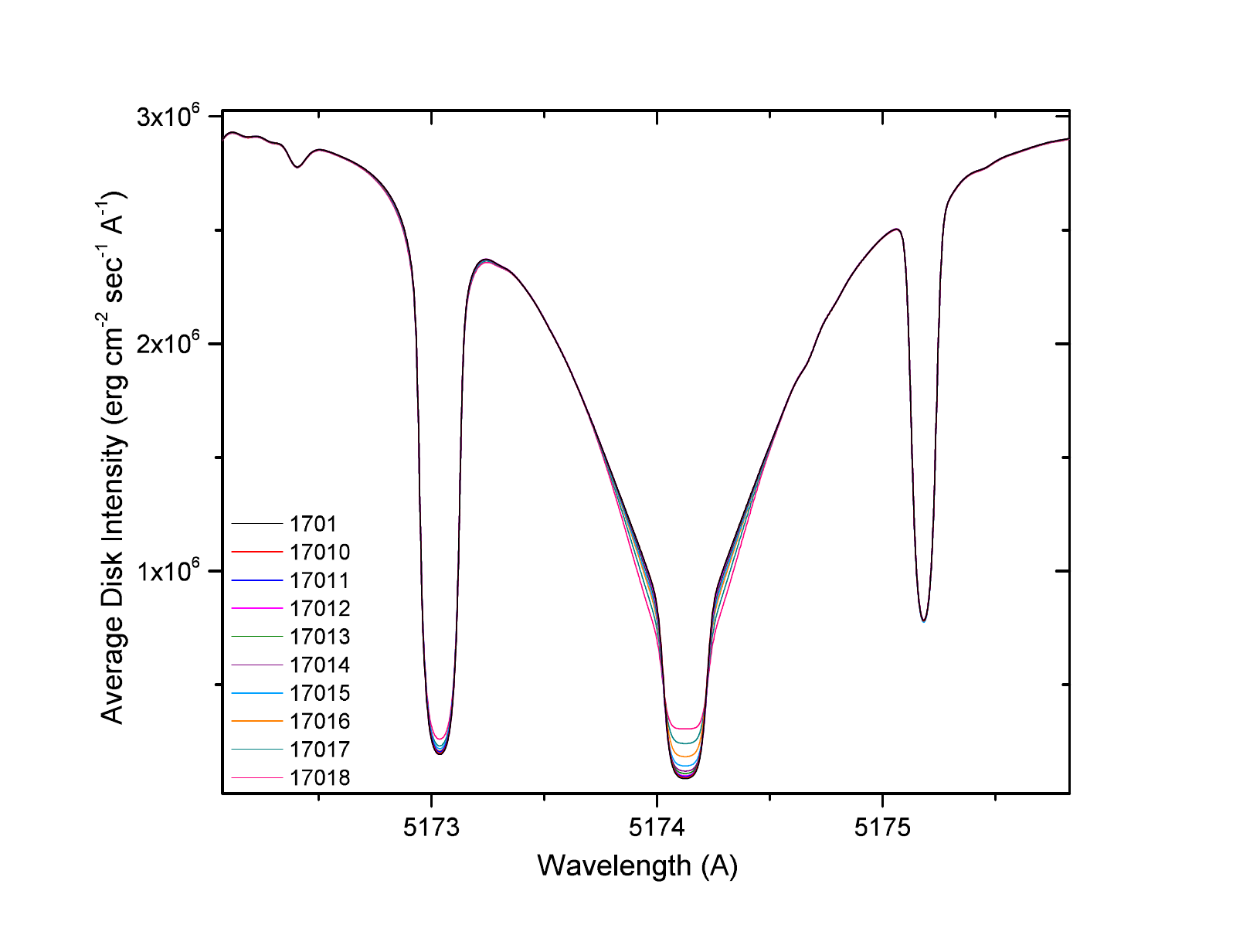}}
\caption{Mg I b2 line profiles (vacuum wavelength) calculated with the set of models. The core as well as the width of the line change with increasing chromospheric activity.}
\label{fig:mg1b2}
\end{figure}

The quiet solar model used here, model 1701, is the same as model 1401 from \cite{fon15} except for recalculating using this new Fe~I atomic model.  Model 1701 is therefore used as the starting point for constructing the set of models with varying activity levels as shown in Figure \ref{fig:modelsyca2} and described in Table \ref{tab:models}.

\section{The changes in the spectra with chromospheric activity} \label{sec:diff}
As a first step in our analysis, we examined the synthetic spectra to compare the characteristics of the lines produced by each model. As expected, the core depth of some lines changed. Several lines also broadened with increasing mechanical heating, as seen with the Mg I b triplet. One of the lines of this triplet (the b2 line) is shown in Figure \ref{fig:mg1b2}. 

\begin{figure*}[t]
    \gridline{
        \includegraphics[trim=1.7cm 1cm 3cm 1cm, clip, width=0.33\textwidth]{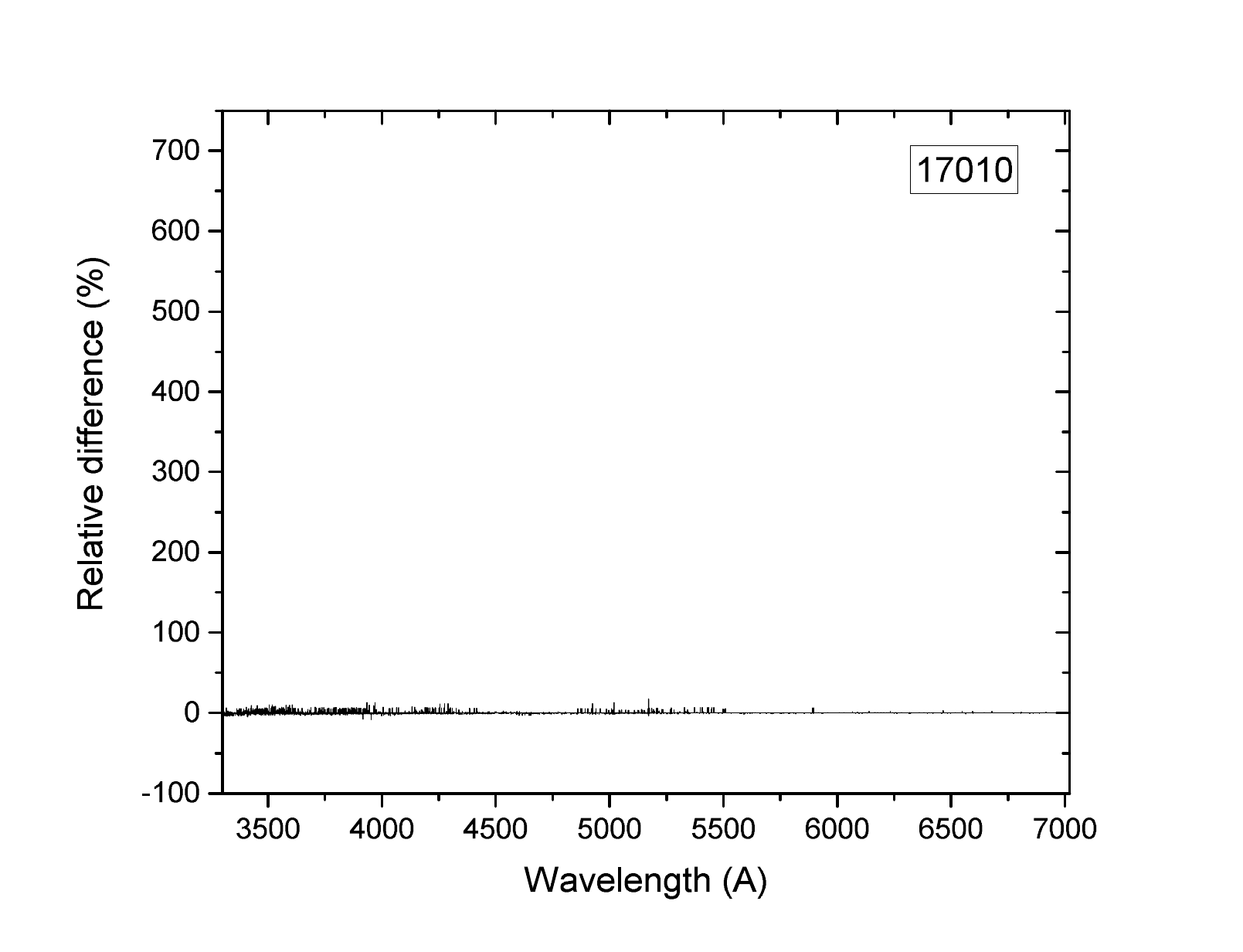}
        \includegraphics[trim=1.7cm 1cm 3cm 1cm, clip, width=0.33\textwidth]{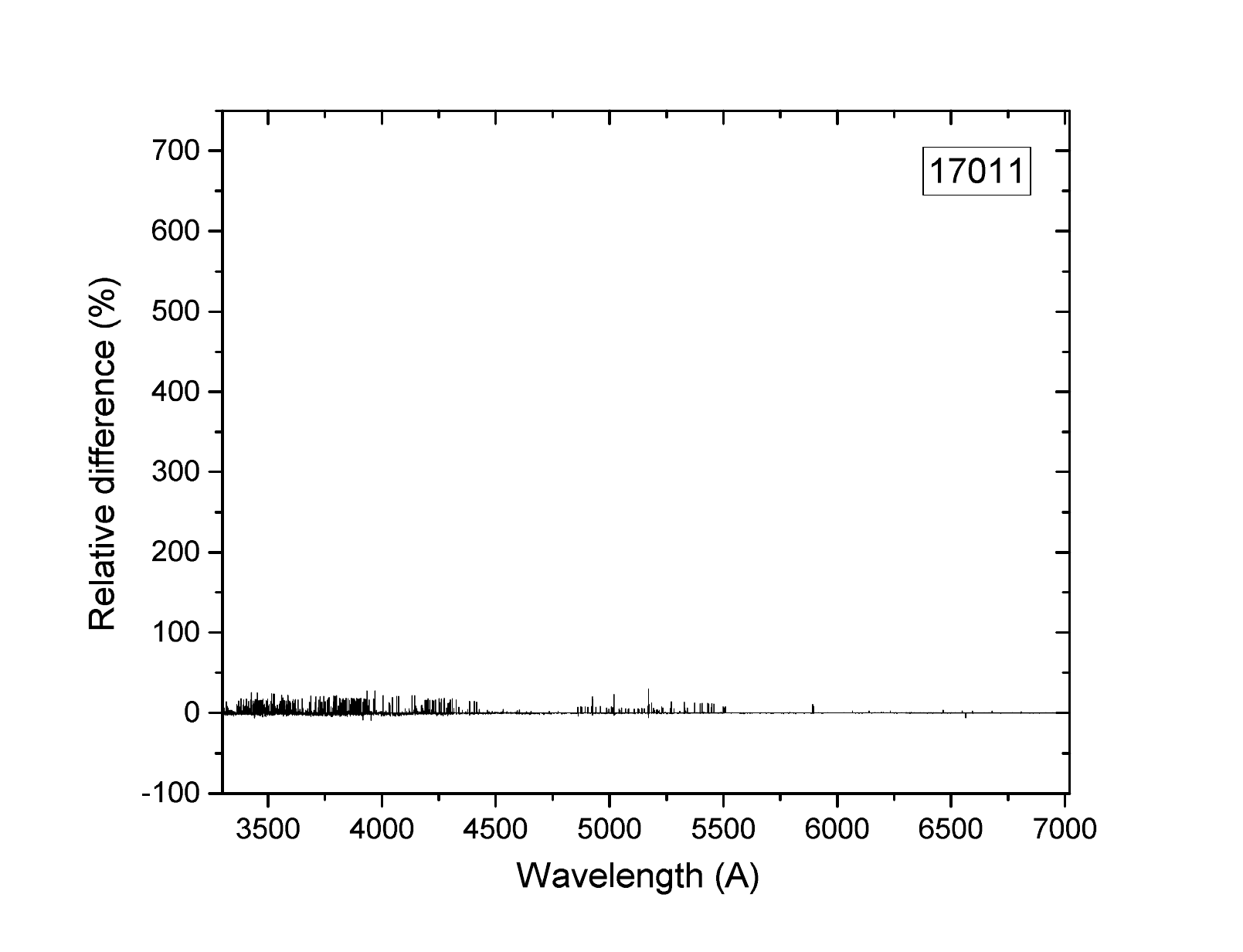}
        \includegraphics[trim=1.7cm 1cm 3cm 1cm, clip, width=0.33\textwidth]{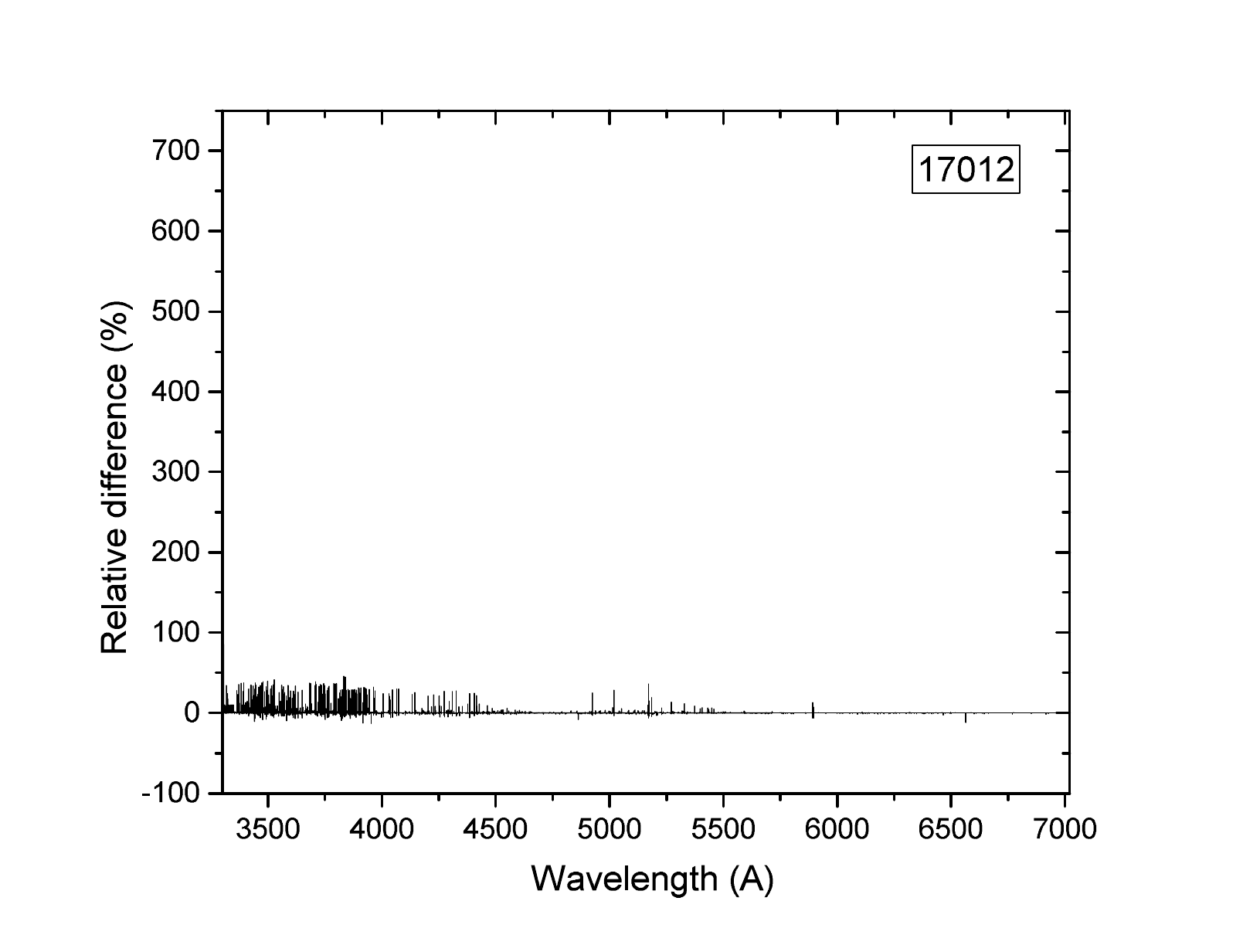}
    }
    \gridline{
        \includegraphics[trim=1.7cm 1cm 3cm 1cm, clip, width=0.33\textwidth]{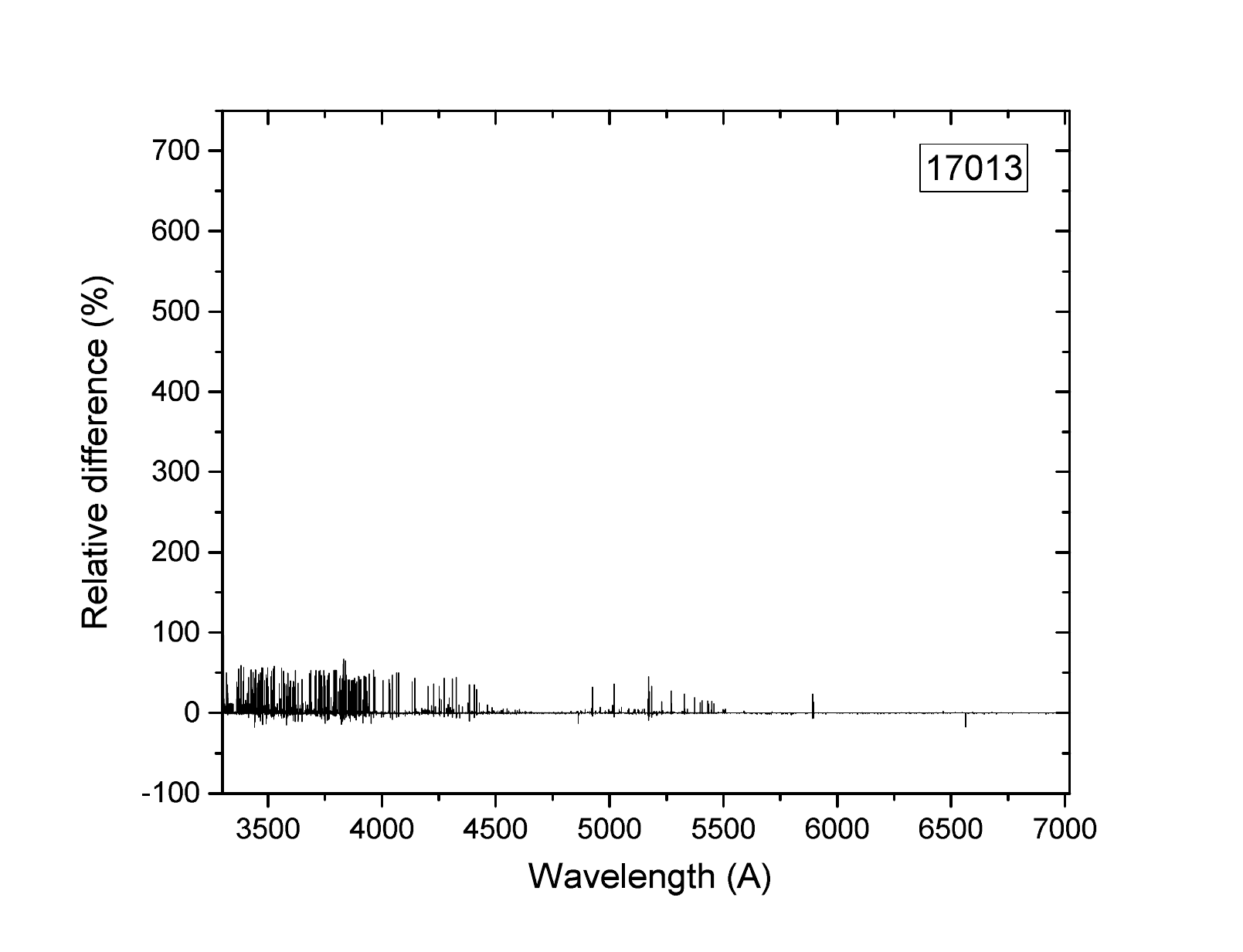}
        \includegraphics[trim=1.7cm 1cm 3cm 1cm, clip, width=0.33\textwidth]{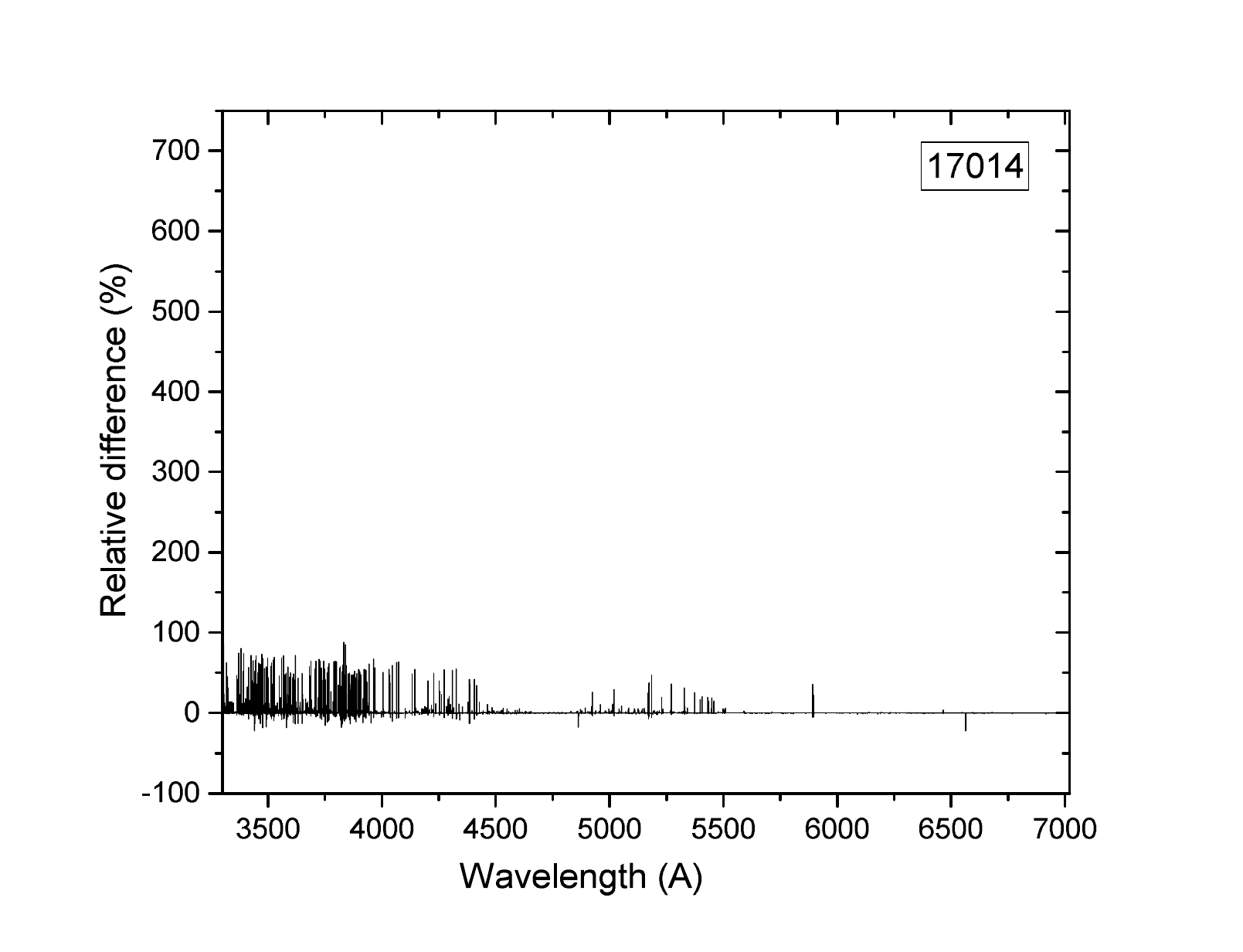}
        \includegraphics[trim=1.7cm 1cm 3cm 1cm, clip, width=0.33\textwidth]{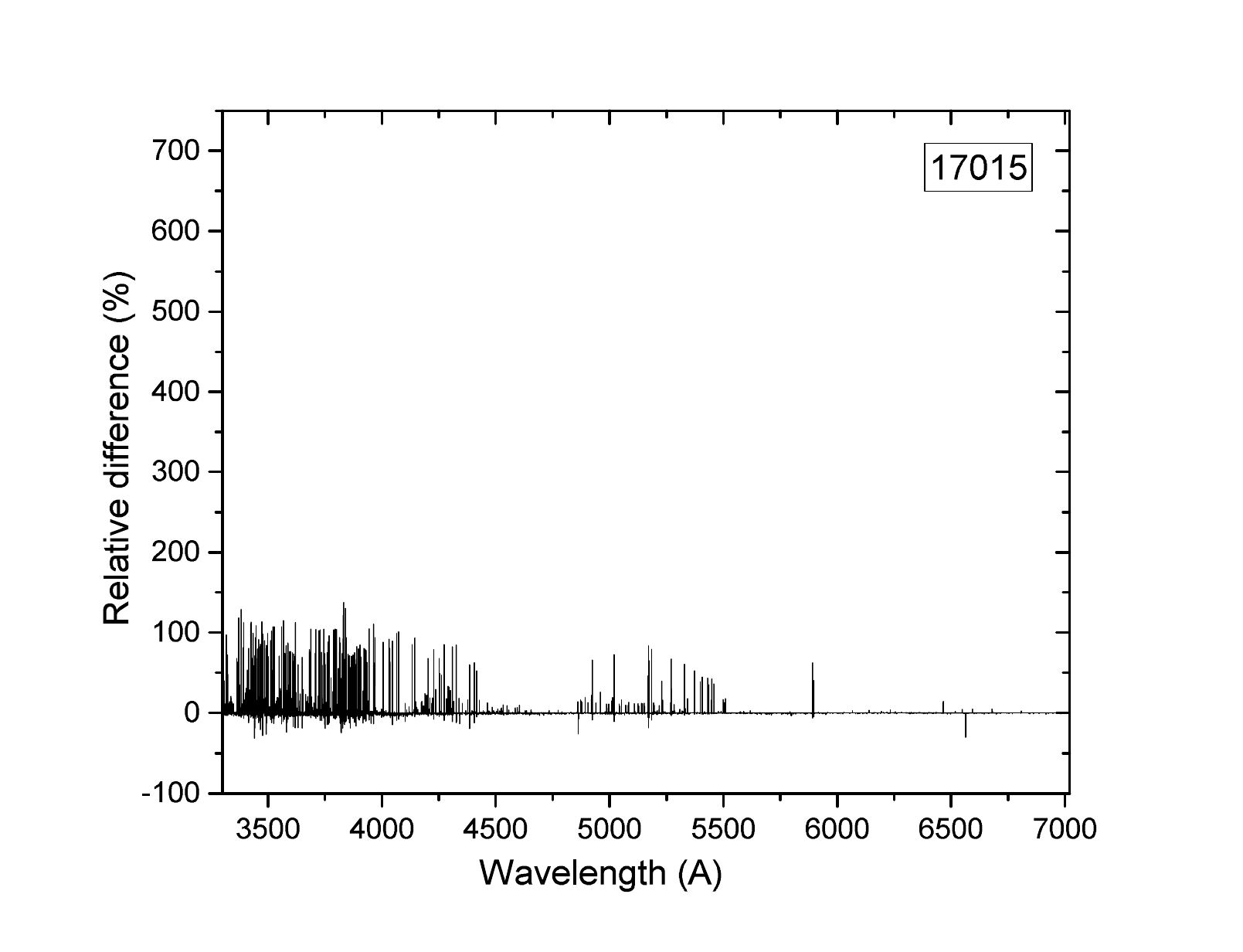}
    }
    \gridline{
        \includegraphics[trim=1.7cm 1cm 3cm 1cm, clip, width=0.33\textwidth]{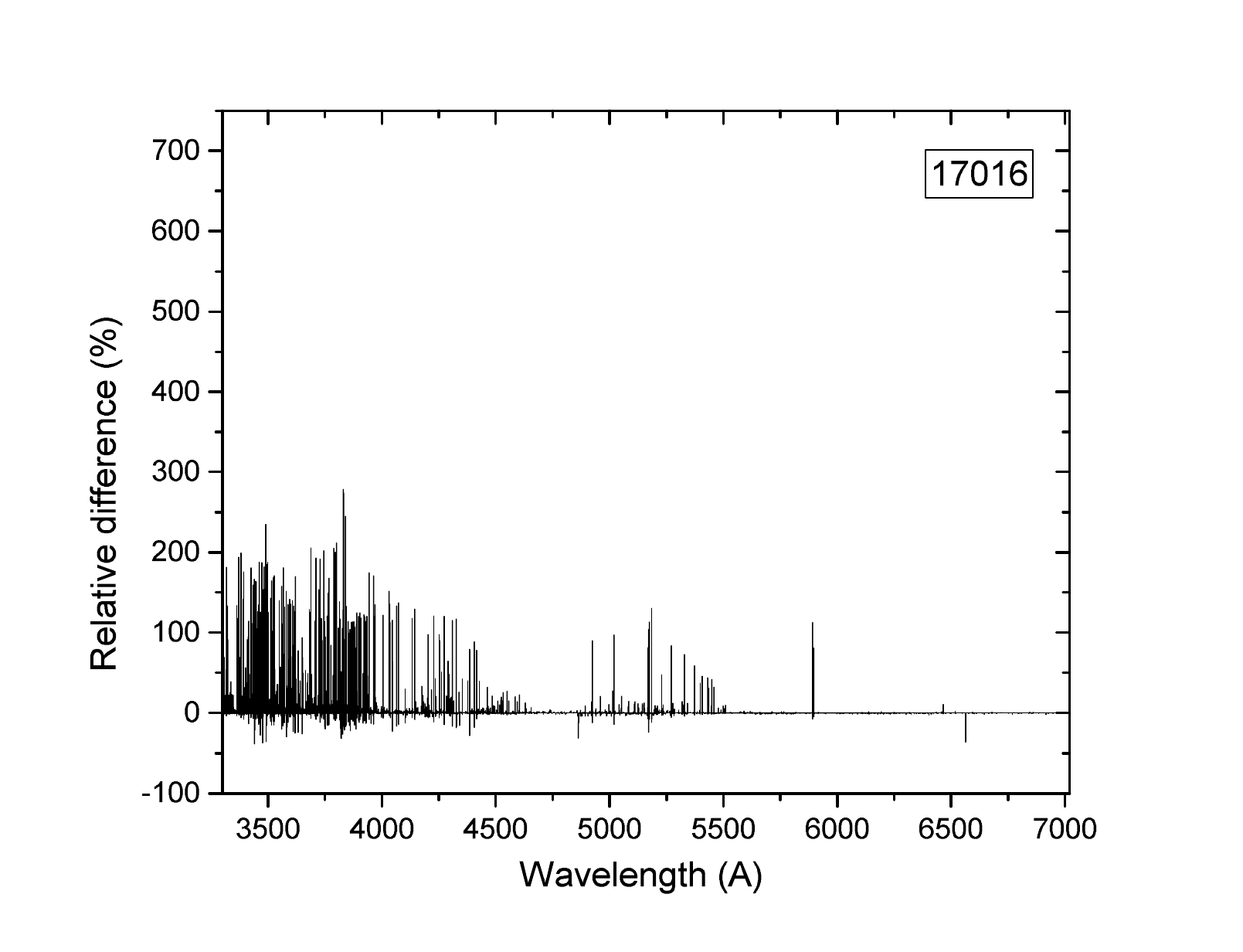}
        \includegraphics[trim=1.7cm 1cm 3cm 1cm, clip, width=0.33\textwidth]{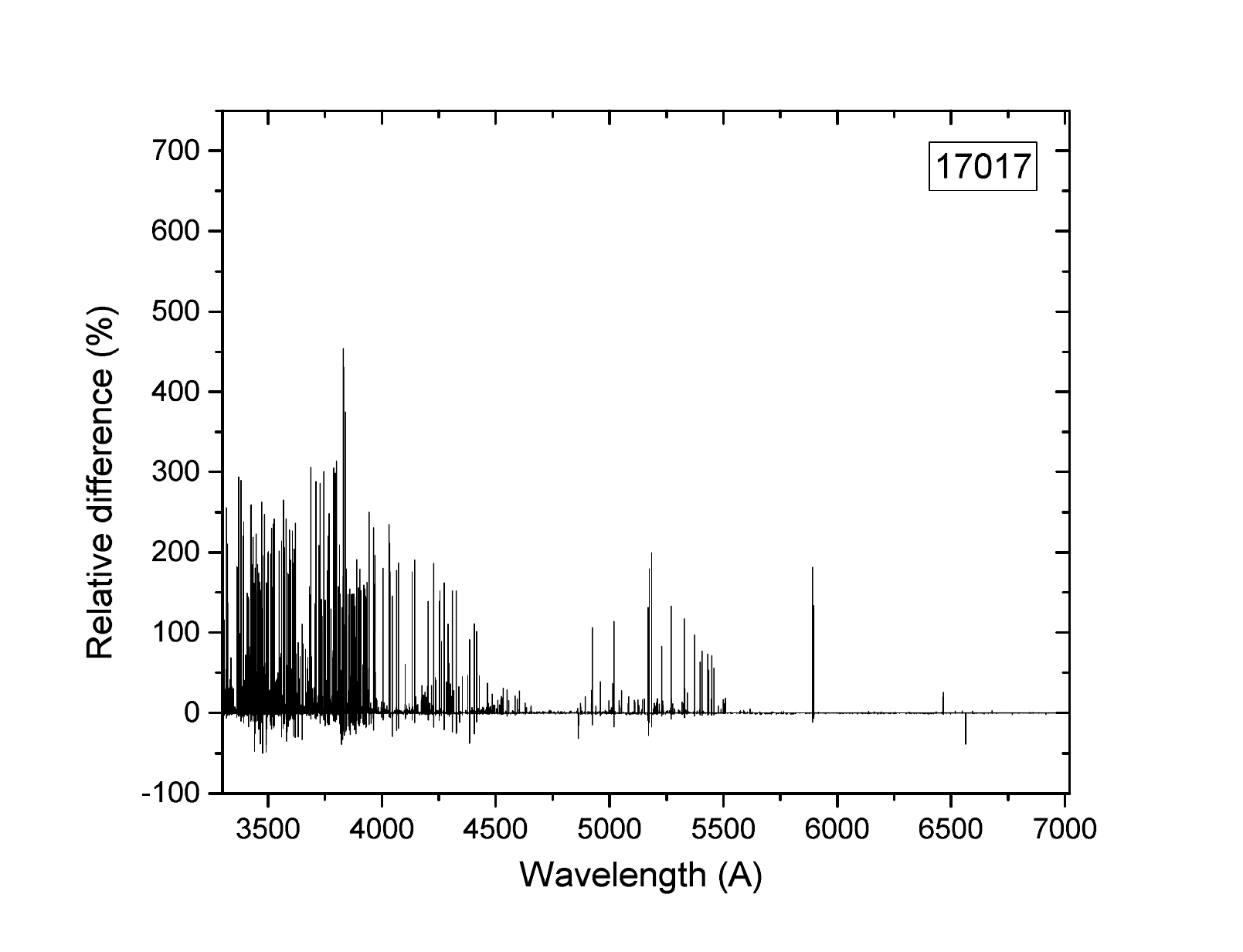}
        \includegraphics[trim=1.7cm 1cm 3cm 1cm, clip, width=0.33\textwidth]{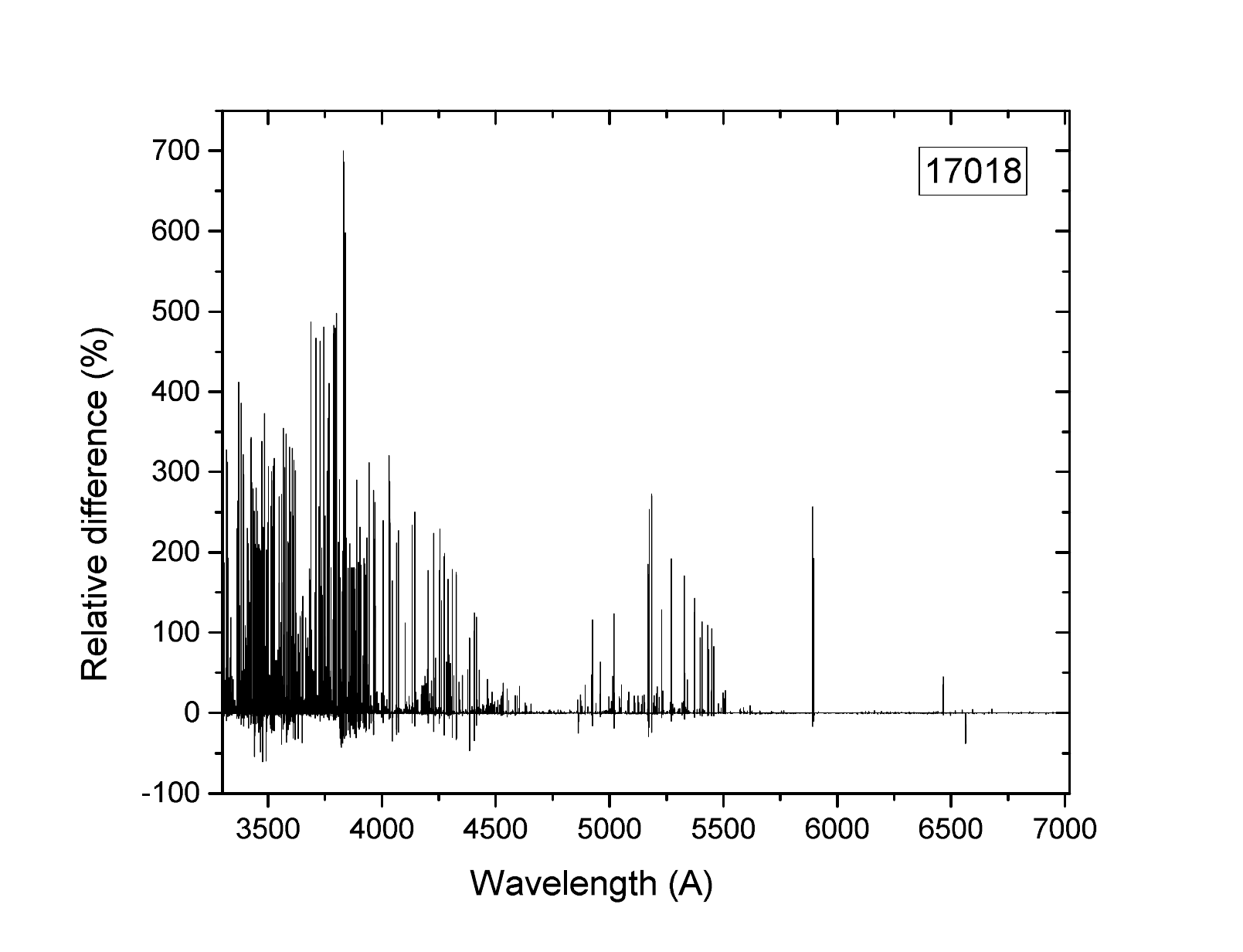}
    }
    \caption{Ratio of lines in each synthetic spectrum relative to those in the lowest-activity model 1701. This is measured by the Relative Difference parameter (RD, in percentage), which represents the relative change in the spectral flux due to stellar activity (vacuum wavelength).}
    \label{fig:diff}
\end{figure*}

To quantify the difference between models due to increasing mechanical heating, we used the Relative Difference (RD) for the ratio of lines in each synthetic spectrum relative to those in the lowest-activity model 1701, expressed as a percentage (i.e., following Eq.\ref{ec:rd_ca} but substituting the integrated flux of the Ca II K line with the flux in the synthetic spectrum). The RD parameter for each wavelength in our synthetic spectrum was used to plot Figure \ref{fig:diff}.
This figure illustrates spectral differences due to chromospheric activity, starting with the upper panel, which shows the RD between models 17010 and 1701, through to the last panel, which shows the RD between models 17018 and 1701. 

In a subsequent step, we cross-matched the atomic line transitions considered in our calculation with the wavelengths in the synthetic spectrum to determine their RD parameter. Since the synthetic spectra were computed with a wavelength step size of $10^{-5}$\AA, we were able to precisely identify the atomic line transition wavelengths in our models taken from the NIST database (see Sections \ref{sec:models} and \ref{sec:atomic}). For line in absorption with respect to the neighboring continuum, the RD corresponds to the change in flux at the very line-bottom. If lines decrease in depth (i.e., the absorption profiles become shallower), the RD thus becomes positive. 

The spectral range shown in Figure \ref{fig:diff} contains 6762 atomic lines, spanning elements from H to Ni in their first ionization states (I, II, and III), with 1715 of these lines being Fe I lines. Table 2 (full version available online) provides the calculated line list and the RD for each model for neutral and the first two ionized atomic species. The first three columns display the parameters of each line—wavelength, atomic number, and ion charge, while the rest of the columns present the RD of the specific line calculated by each model.

These plots in Figure \ref{fig:diff} highlight specific regions where the generated spectra change the most due to chromospheric heating. The most affected wavelength ranges are approximately from 3300 to 4400 \AA~and from 5250 to 5500 \AA. In the last panel, which represents the most active model and where the temperature in the chromospheric plateau differs by 1200 K, the difference in some of the lines in the first wavelength range can reach approximately 700 \%, while in the second range the RD can reach around 300 \%. From 5500 \AA~to the end of the studied range, the only regions significantly impacted by the temperature increase are the Na I D doublet and H$\alpha$ line profiles. In the same spectral range, starting from the panel with model 17015 to the end, an unexpected Ca I line (6464.353 \AA) begins to respond significantly to the temperature increase. This line highlights the fact that certain spectral lines can start changing at a specific level of activity if the temperature increase is sufficient to enhance the atomic processes driving the change.

Another interesting observation is that most absorption lines decrease in depth related to the neighboring continuum as chromospheric activity increases, such as the cores of the Mg I b2 and the line near 5173 \AA~shown in Figure \ref{fig:mg1b2}, which corresponds to a positive RD in Figure \ref{fig:diff} or Table 2. However, there are some lines that deepen with increasing activity, such as H$\alpha$, resulting in negative RDs.

It is important to note that the RDs presented in Figure \ref{fig:diff} and Table 2 are calculated without reducing the spectral resolution to that of an actual spectrograph. This work focuses on analyzing the variations in line formation caused by chromospheric activity. Consequently, this assumption may result in real stellar spectra being unable to distinguish differences in the line profiles for smaller RDs, depending on their resolution. Actual values for lines broadened by stellar rotation or finite spectrograph resolution will therefore be somewhat different.

\begin{figure}[t]
\centering
\resizebox{\hsize}{!}{\includegraphics[trim=2cm 1cm 1cm 2cm, clip, width=1.0\textwidth]{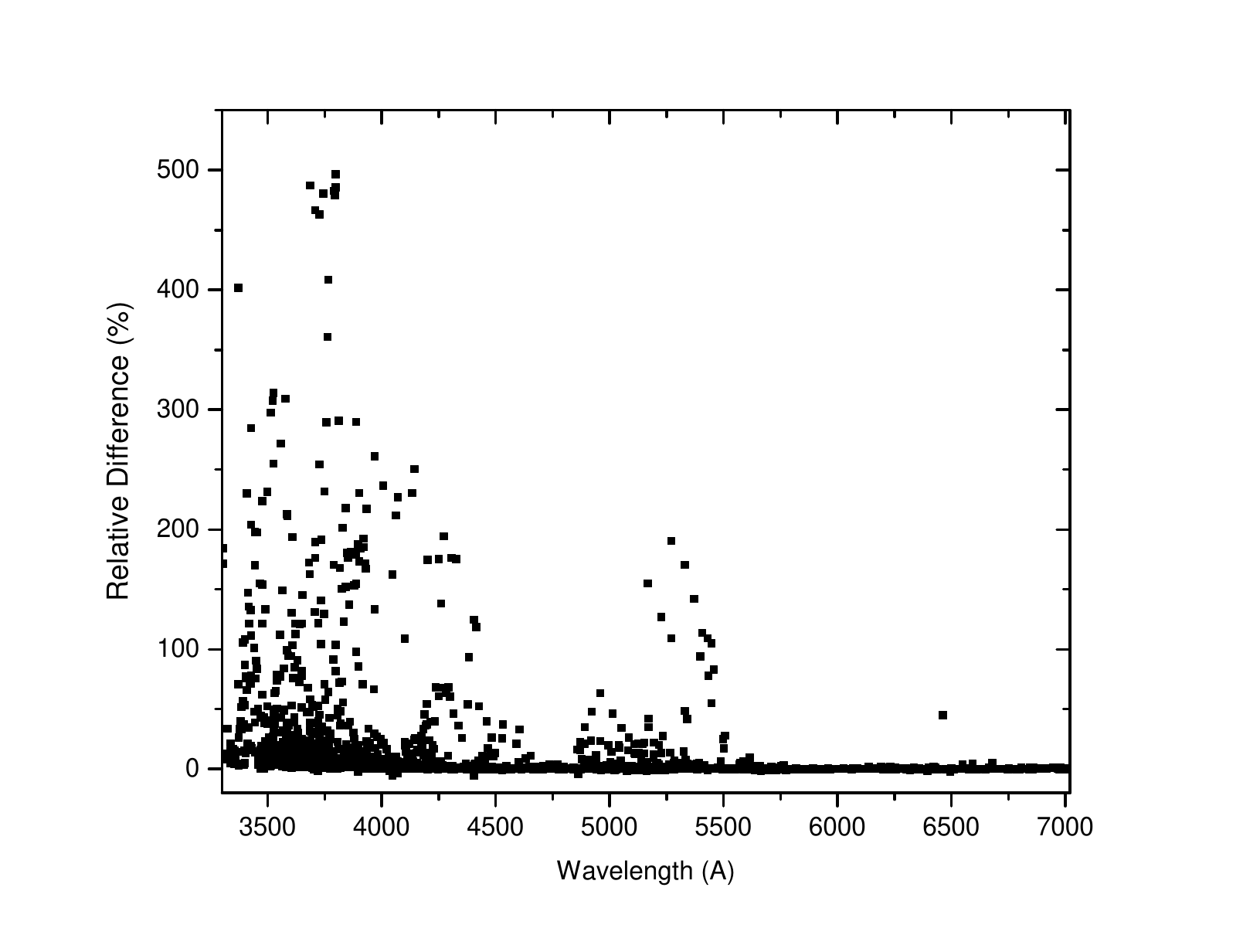}}
\caption{Ratio of Fe I lines calculated using the highest-activity model (17018) relative to those in the lowest-activity model 1701. This is measured by the Relative Difference parameter (RD, in percentage), which represents the relative change in the spectral flux due to stellar activity (vacuum wavelength) (taken from \protect\cite{viey24}).}
\label{fig:onlyFeI}
\end{figure}

The RD of the hottest model in our set (17018) considering only the Fe I lines was previously presented in Figure 3 of \cite{viey24}, and is reproduced here in Figure \ref{fig:onlyFeI}. Comparing this figure with the last panel of Figure \ref{fig:diff}, this shows a similar trend of changes with chromospheric activity (for more details see that work).

\begin{deluxetable*}{cccccccccccc}
\tablecaption{First ten lines of the calculated line list for neutral and first two ionized atomic species, the rest of the lines are available as online material in the machine-readable format. The first three columns are the wavelength (vacuum), atomic number, and ion charge of the transition.  The rest of the columns contains the line's respective Relative Difference (RD) in percentage for the lines calculated with each model relative to the lowest-activity model 1701. The RD parameter represents the change in the lines due to stellar activity.}
\tablewidth{0pt}
\tablehead{
\colhead{ Wavelength} & \colhead{Atomic}  & \colhead{Ion} & \colhead{ 17010}& \colhead{ 17011}& \colhead{ 17012}& \colhead{17013}& \colhead{ 17014}& \colhead{ 17015}& \colhead{ 17016}& \colhead{ 17017}& \colhead{17018}\\
 \colhead{ (\AA)} & \colhead{ Number}  & \colhead{Charge} & \colhead{(\%)}& \colhead{(\%)}& \colhead{(\%)}& \colhead{(\%)} & \colhead{(\%)} & \colhead{(\%)}& \colhead{(\%)}& \colhead{(\%)}
 & \colhead{(\%)}}
\decimalcolnumbers
\startdata
  3300.003  & 12 & 2 & 0.002& -0.007&-0.024&-0.027&-0.062&-0.071&-0.170&-0.253&-0.344 \\
  3300.035 & 23 & 0 & 0.002&-0.006&-0.019&-0.028&-0.058&-0.064&-0.166&-0.250&-0.329 \\
  3300.252 & 10 & 2 &0.003&-0.005&-0.025&-0.026&-0.061&-0.070&-0.170&-0.252&-0.343\\
  3300.337 & 8 & 2 & 0.004&-0.002&-0.177&-0.019&-0.053&-0.064&-0.152&-0.223&-0.285\\
  3301.465 & 10 & 2 &0.003&-0.004&-0.024&-0.026&-0.060&-0.071&-0.170&-0.252&-0.342\\
  3301.909 & 25 & 0 &0.002&-0.006&-0.025&-0.027&-0.062&-0.073&-0.171&-0.253&-0.343\\
  3302.362 & 8 & 1 & 0.003&-0.004&-0.025&-0.028&-0.062&-0.072&-0.172&-0.255&-0.346\\
  3302.624 & 22 & 1 & -0.300&0.180&-0.316&-0.312&0.008&-1.544&-0.294&-0.473&1.975\\
  3303.319 & 11 &0 & 3.001&2.049&-4.459&-5.808&-3.271&-3.961&-0.063&3.675&11.652\\
  3303.808 & 26 & 1 & 2.692&8.415&25.766&26.864&24.864&39.924&156.785&159.655&162.290\\
\enddata
\label{tab:diff}
\end{deluxetable*}

\subsection{Changes in the Contribution Function} \label{sub:contrib}
To disentangle the behavior of the spectral lines with increasing chromospheric activity, 
we computed the total contribution function ($f_c$) as described by \cite{fon07} at frequency $\nu$, 
\begin{equation}
\mbox{$f_c=(c^2/2h\nu) \, \varepsilon_\nu \, e^{-\tau_\nu}$}
\label{ec:cf}
\end{equation}

where $\varepsilon_\nu$ is the total emissivity and $\tau_\nu$ is the total optical depth at that frequency. 

The $f_c$ allows us to identify the formation region of a spectral line in the stellar atmosphere by locating the peaks of maximum contribution, as done in \cite{per23}. Although we are capable of calculating the $f_c$ of each line in our synthetic spectra, we focused our computation on the Fe I spectral lines. This procedure, however, is applicable to any other species with lines in the spectra. Moreover, since the most significant spectral changes occur in the hottest model in our set, we compared the $f_c$ of this model (17018) with that of the initial model, 1701. 

\begin{figure*}[t]
    \gridline{
        {(a)}{\includegraphics[trim=1.5cm 1.4cm 3cm 2cm, clip, width=0.3\textwidth]{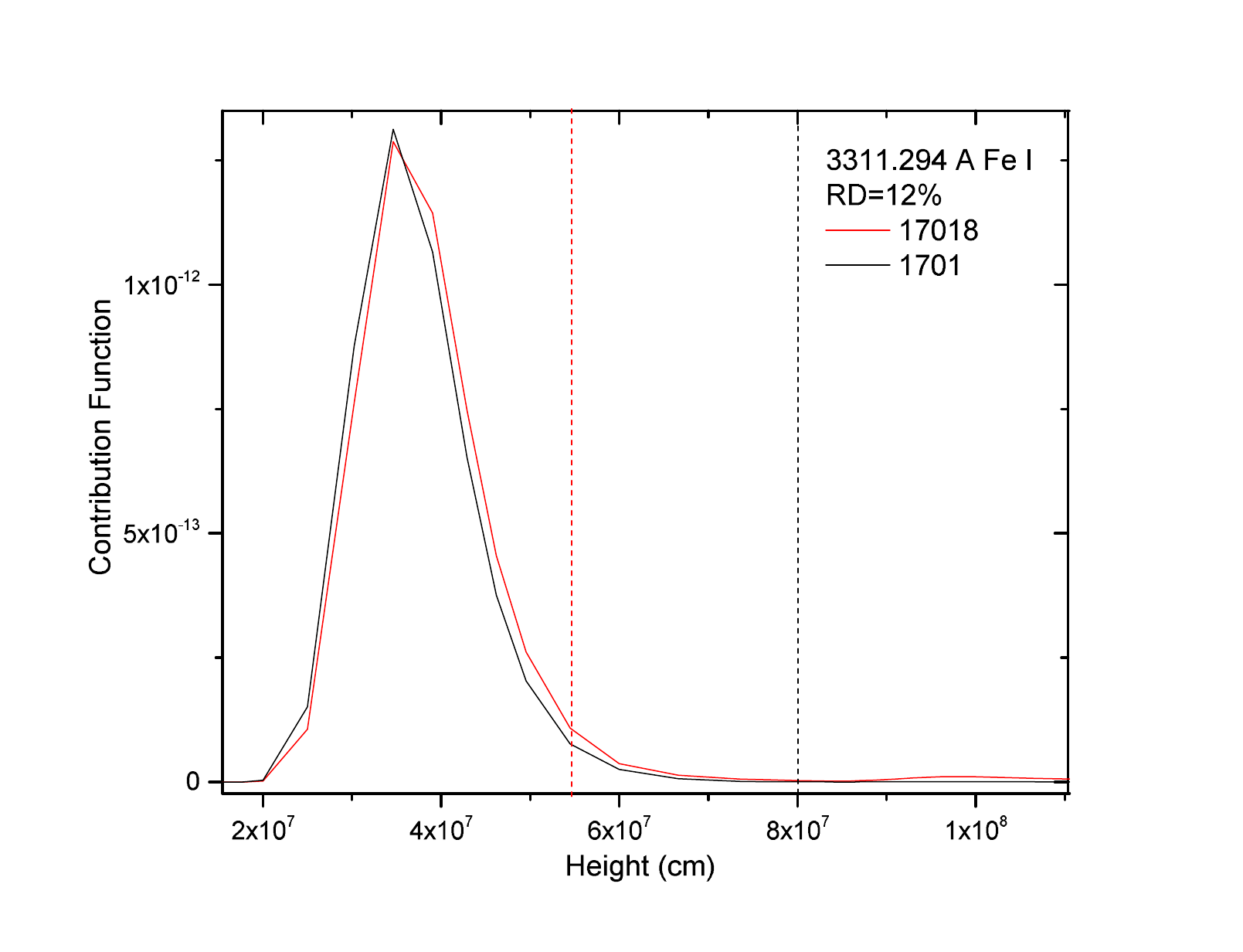}}
      {(b)}{\includegraphics[trim=1.5cm 1.4cm 3cm 2cm, clip, width=0.3\textwidth]{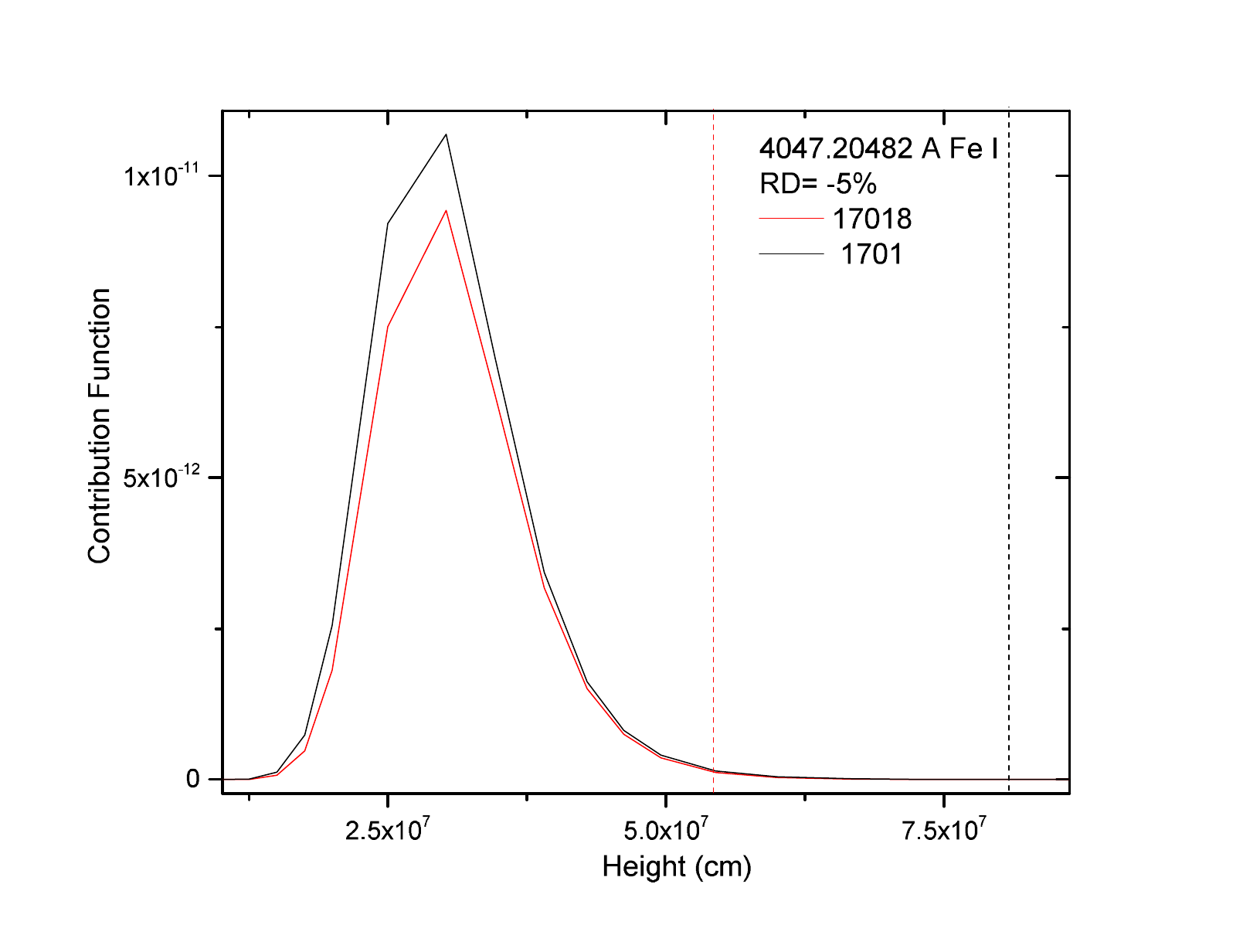}}
      {(c)}{\includegraphics[trim=1.5cm 1.4cm 3cm 2cm, clip, width=0.3\textwidth]{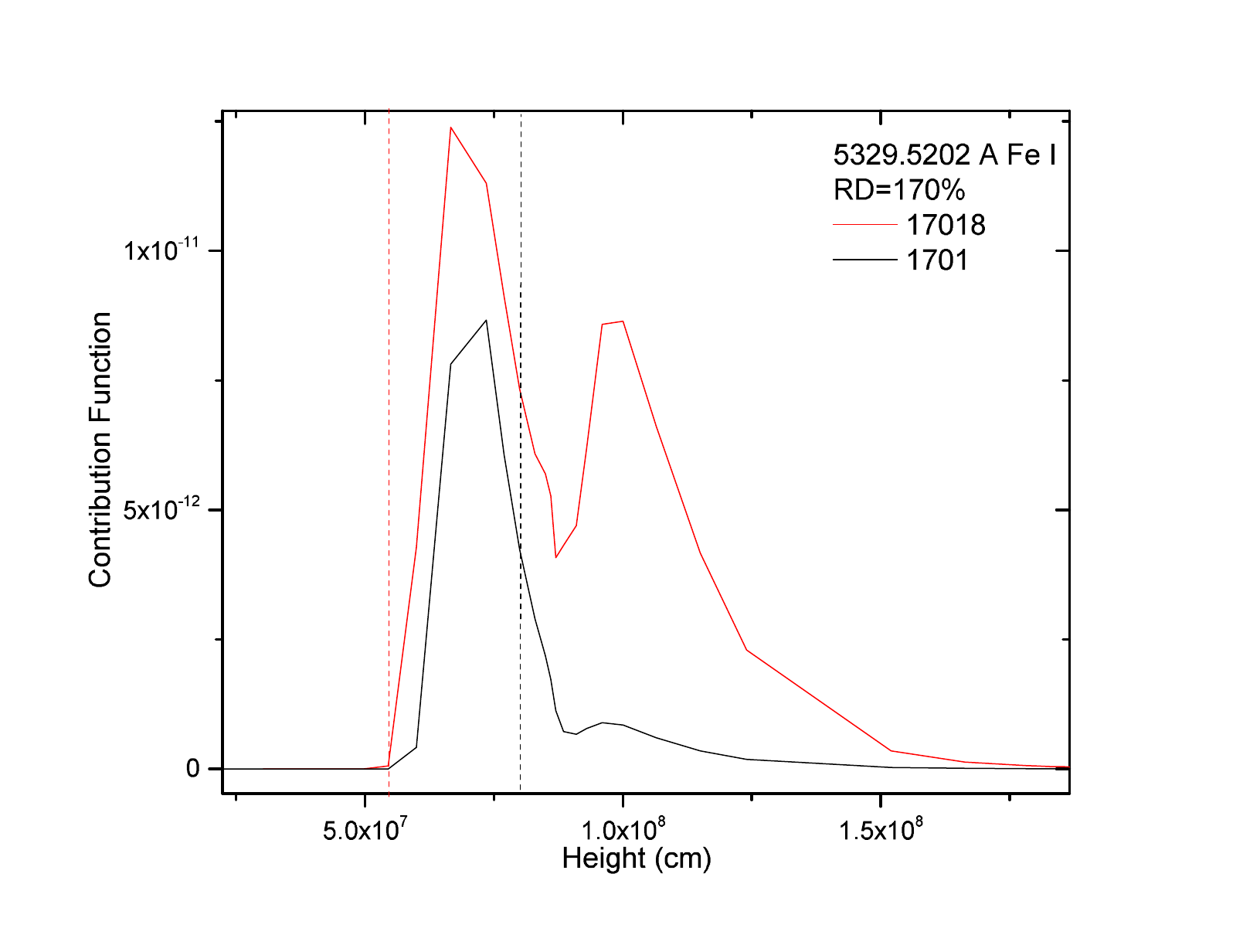}}
    }
    \caption{Contribution function, calculated using Eq. \ref{ec:cf}, for a selection of Fe I lines calculated with 1701 (lowest-activity model, black) and 17018 (highest-activity model, red). Also displayed are their corresponding RD between these two models, and a dashed line indicating the location of the temperature minimum for each model.}
    \label{fig:fc}
\end{figure*}

Figure \ref{fig:fc} illustrates the $f_c$ for three Fe I lines calculated using models 1701 (black) and 17018 (red), along with their corresponding RD values. The figure also shows the location of the temperature minimum for each model (dashed lines). The line profiles in panels (a) and (b) exhibit lower RD values, while the profile in panel (c) shows higher RD value. In panel (c) and, to a lesser extent in panel (a), the most active model exhibits not only an increase in the main peak of $f_c$ but also the appearance of a secondary peak further out in the atmosphere, contributing to the line formation. The increase in the main peak and the emergence of a secondary peak appear to be the primary factors driving the observed changes in the lines.

\begin{figure*}[t]
\centering
\resizebox{\hsize}{!}{\includegraphics[trim=2cm 1cm 2cm 2cm, clip, width=1.0\textwidth]{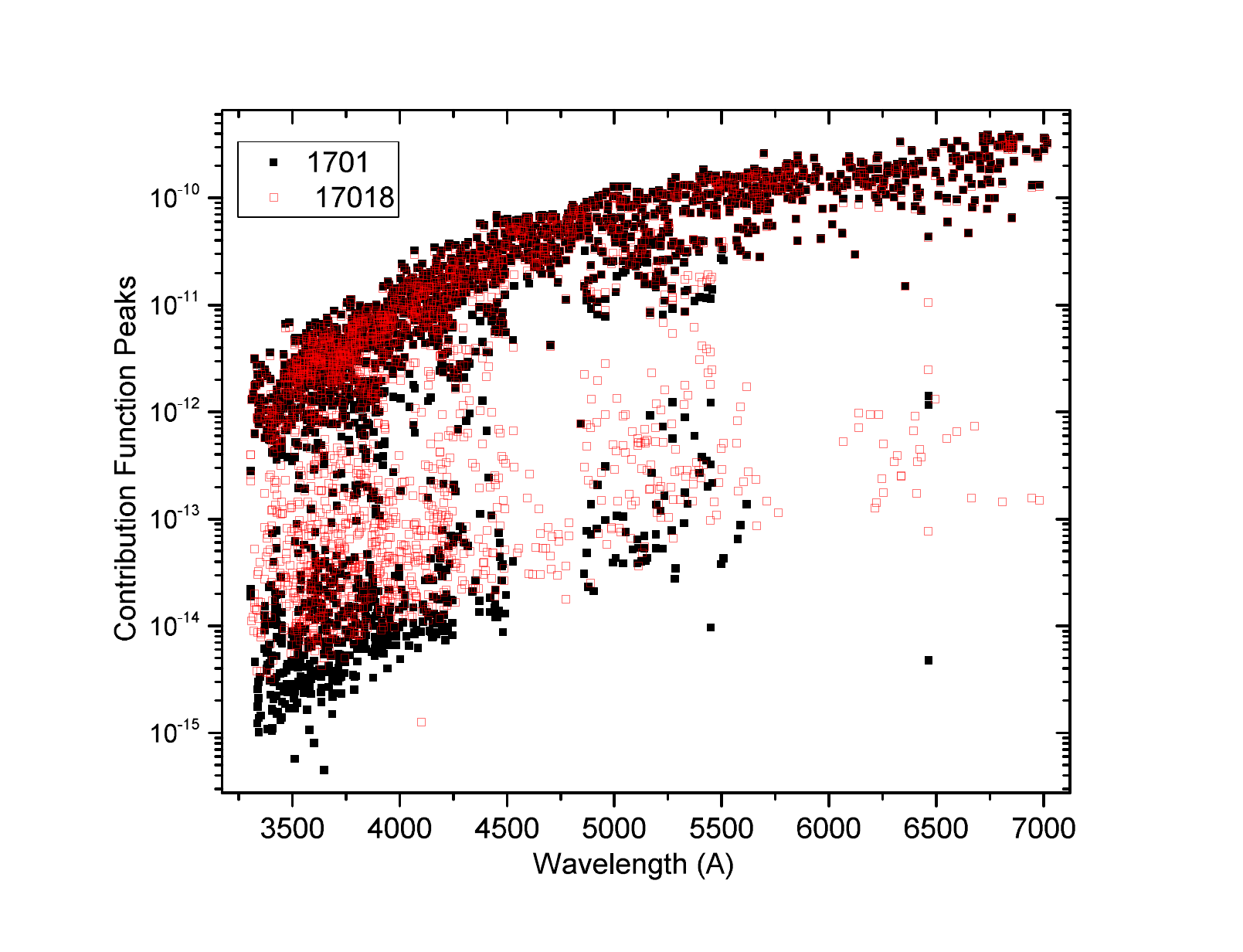}}
\caption{ Filtered peaks for the Contribution Function, calculated using Eq. \ref{ec:cf}, of Fe I lines calculated from the lowest-activity model 1701 (black) and the highest-activity model 17018 (red) as a function of wavelength (vacuum).}
\label{fig:fcvswave}
\end{figure*}

To investigate further, we calculated the peaks of the $f_c$ for each Fe I line for models 1701 and 17018. The results is shown in Figure \ref{fig:fcvswave} where the peaks for all lines calculated from models 1701 (black) and 17018 (red) are plotted as a function of wavelength. Since multiple peaks can occur at a given wavelength, we applied a criterion to retain only those peaks that are at most three orders of magnitude lower than the main peak for each line. Peaks below this threshold are considered negligible for line formation. The retained peaks are referred to as “Filtered Peaks” throughout the rest of the paper.

When comparing this figure with Figure \ref{fig:onlyFeI}, it is possible to approximately identify the changes in the lines shown in the latter. There are several structures, with a wide band at the top for both models, corresponding to the lower band near a RD of 0~\% in Figure \ref{fig:onlyFeI}. Below this main band, secondary peaks are observed, which, in most cases, shift to higher values of $f_c$ from model 1701 to model 17018. Additionally, several new secondary peaks appear exclusively in model 17018. 
The upward shift and the appearance of new secondary peaks could explain the changes observed in the RD of the Fe I lines. It is important to note that the wavelength ranges where these secondary peaks appear or shift upward in the most active model, 17018, correspond with the spectral ranges that show changes in  Figure \ref{fig:onlyFeI}, reflecting an enhanced chromospheric contribution to line formation as chromospheric temperature increases.

\begin{figure}[t]
\centering
\resizebox{\hsize}{!}{\includegraphics[trim=2cm 1cm 1cm 1cm, clip, width=1.0\textwidth]{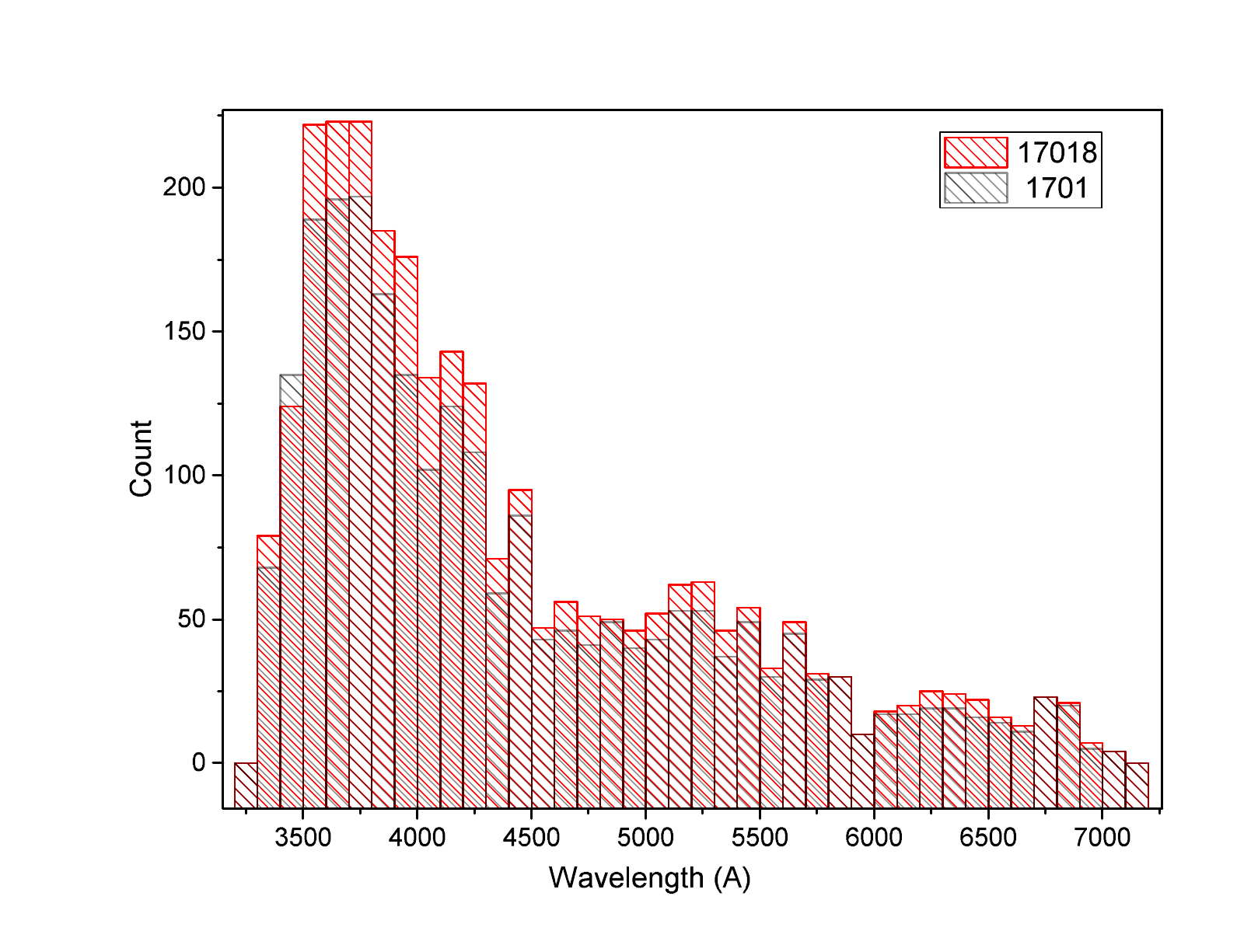}}
\caption{ Number of filtered peaks for the Contribution Function, calculated using Eq.\ref {ec:cf}, of Fe I lines calculated from the lowest-activity model 1701 (black) and the highest-activity model 17018 (red) as a function of wavelength.}
\label{fig:histogram_peaks}
\end{figure}

Figure \ref{fig:histogram_peaks} compares the number of $f_c$ filtered peaks per 100 \AA -wide bin between both models, 1701 and 17018. In most Fe I lines, an additional secondary peak contributing to line formation appears within the spectral range where the lines exhibit significant changes with activity in model 17018.

\begin{figure}[t]
\centering
\resizebox{\hsize}{!}{\includegraphics[trim=2cm 1cm 2cm 2cm, clip, width=1.0\textwidth]{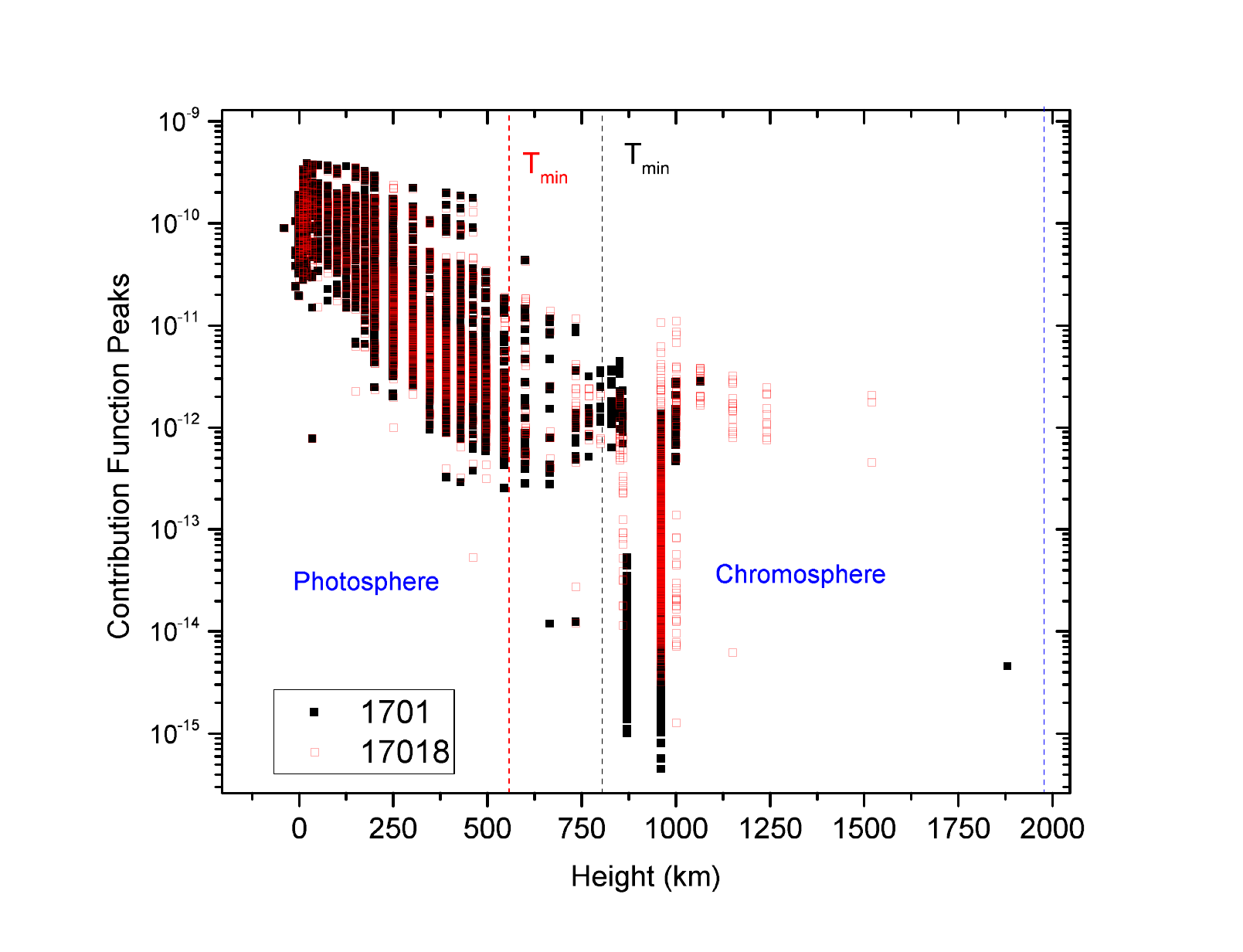}}
\caption{ Location in the stellar atmosphere of the filtered peaks for the Contribution Function, calculated using Eq. \ref{ec:cf}, of Fe I lines calculated from the lowest-activity model 1701 (black) and the highest-activity model 17018 (red).}
\label{fig:cfvsh}
\end{figure}

Figure \ref{fig:cfvsh} illustrates the location of the $f_c$ filtered peaks in the stellar atmosphere for both models (1701 in black, 17018 in red). The vertical dashed lines indicate the location of the temperature minimum for each model, marking the boundary between the photosphere and the chromosphere. The blue dashed line indicates the beginning of the transition region. As shown in this figure, most of the new $f_c$ peaks in model 17018 appear in the chromosphere.  It is noteworthy that the temperature minimum shifts inward from model 17014 to model 17018 (see Table \ref{tab:models}) in line with the results of \cite{viey05}. This indicates that the first chromospheric rise for model 17018 occurs at heights corresponding to the photosphere in the thermal structure of model 1701.

It is important to note that in models including a chromosphere, like ours, the same temperature may be repeated at different atmospheric heights that correspond to the photosphere or the chromosphere (see Figure \ref{fig:model_ca2}). This implies that the temperature alone is not a reliable parameter for identifying the formation region of a line, as it can be when using photospheric models only, as in \cite{almou22}. In such cases, the peaks of the contribution function are more accurately determined by considering both the temperature and the pressure (or the height) in the atmosphere where they originate. 

The results presented in this section support the hypothesis that chromospheric heating caused by magnetic activity can alter the line profiles of transitions whose contribution function either increases, decreases at the same location in the atmosphere, or introduces a new contribution farther outward in the atmosphere. 

\subsection{Comparison of our results with observational previous works} \label{sub:compa}
To place an observational constraint on our findings, we cross-referenced our calculated list of line profiles with those previously identified in the literature, as discussed in the Introduction. 

\cite{wise18} presented a list of 43 activity-sensitive lines comparing high resolution observations acquired by HARPS of two dK2 stars, $\epsilon$ Eridani and $\alpha$ Cen B. Of the lines they identified, 17 lines coincide with those in our list. Table \ref{tab:wise} lists the wavelength of these lines, the associated atomic species, and the RD of our most active model, 17018, compared to the base model 1701. For six of these lines, the RD is under 2.1\%, while for the remaining 11 lines the RD exceeds 12\%. A possible explanation for the smaller RD in six lines could be that model 17018 is representative of a star with a lower chromospheric activity than $\epsilon$ Eridani.

\begin{figure}[t]
\centering
\resizebox{\hsize}{!}{\includegraphics[trim=2cm 1cm 2cm 2cm, clip, width=1.0\textwidth]{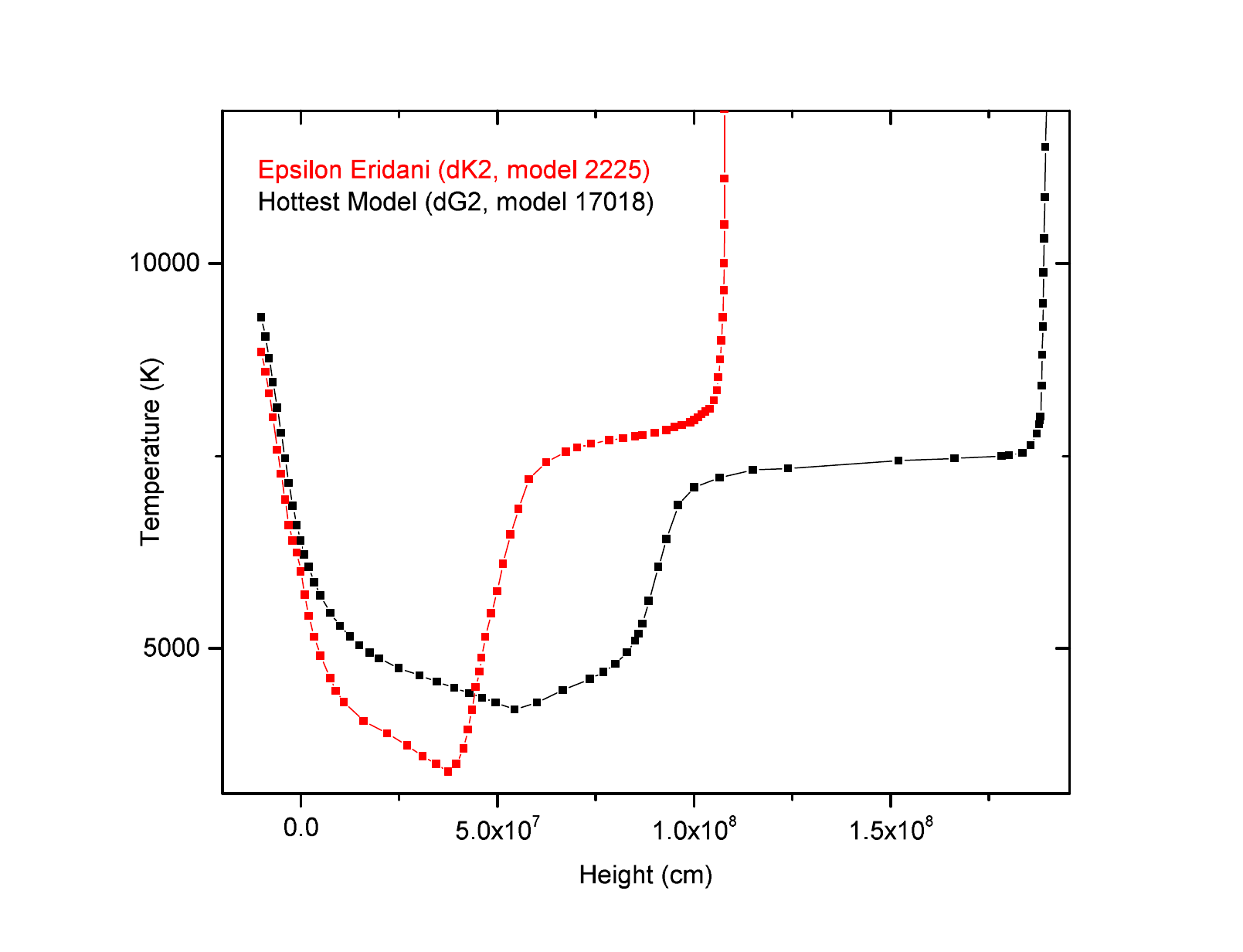}}
\caption{ Comparison between our most active model 17018, and a model that represents $\epsilon$ Eridani built by \protect\cite{viey20}.}
\label{fig:EpsSun}
\end{figure}

Figure~\ref{fig:EpsSun} compares the atmospheric model of our most active model, 17018, with model 2225, which represents the active star $\epsilon$ Eridani built by \cite{viey20}. We observed that from the first chromospheric rise to the onset of the transition region, the chromosphere of model 17018 exhibits a cooler thermal structure than model 2225. This suggests that some lines that showing variation in $\epsilon$ Eridani, an active dK2 star, may not exhibit the same behavior in the most active model of our fictitious dG2 star. This is because those lines form in model 17018 at lower temperatures. Consequently, our most active model represents a star that is less chromospherically active than $\epsilon$ Eridani.
\begin{deluxetable}{lccl}
\tablecaption{List of lines that coincide between this work and \cite{wise18}. The first three columns are the wavelength, atomic number, and ion charge of the transition.  The fourth column contains their respective RD between models 1701 and 17018.\label{tab:wise}}
\tablewidth{0pt}
\tablehead{
\colhead{ Wavelength} & \colhead{Atomic}  & \colhead{Ion} & \colhead{RD}\\
 \colhead{ (\AA)} & \colhead{ Number}  & \colhead{Charge} & \colhead{(\%)
 }}
\decimalcolnumbers
\startdata
4572.3767&12&0&2.04\\
4828.807&23&0&	-0.14\\
5084.755&	26&	0&	26.08\\
5108.8705&	26&	0&	20.66\\
5111.837&	26&	0&12.33\\
5174.1251&	12&	0&	253.50\\
5185.0479&	12&	0&	272.59\\
5251.6703&	26&	0&	1.47\\
5398.6279&	26&	0&	93.76\\
5407.2772&	26&	0&	113.59 \\ 
5421.861 & 25& 0& -0.17\\
5431.2054&	26&	0&	109.18\\
5508.3083&	26&	0&	27.78\\
5897.558&	11&	0&	192.71\\
5897.565&	16&	1&	192.71\\
6232.4455&	26&	0&	1.70\\
\enddata
\end{deluxetable}

\cite{spi20} used a line list consisting of 241 lines of various atomic species including Fe, C, Na, Mg, Al, Si, S, Ca, Sc, Ti, V, Cr, Mn, Co, Ni, Cu, Zn, Y, Zr, and Ba. Analyzing HARPS observations of sun-like stars along with their activity index $R'_{HK}$, they investigated  how equivalent widths (EWs) of these lines vary with chromospheric activity, as detailed in their Table 2. Although some of the atomic species included in \cite{spi20} are not present in our calculations, we found 164 matching lines.

\begin{figure}[h]
\centering
\resizebox{\hsize}{!}{\includegraphics[trim=2cm 1cm 2cm 2cm, clip, width=1.0\textwidth]{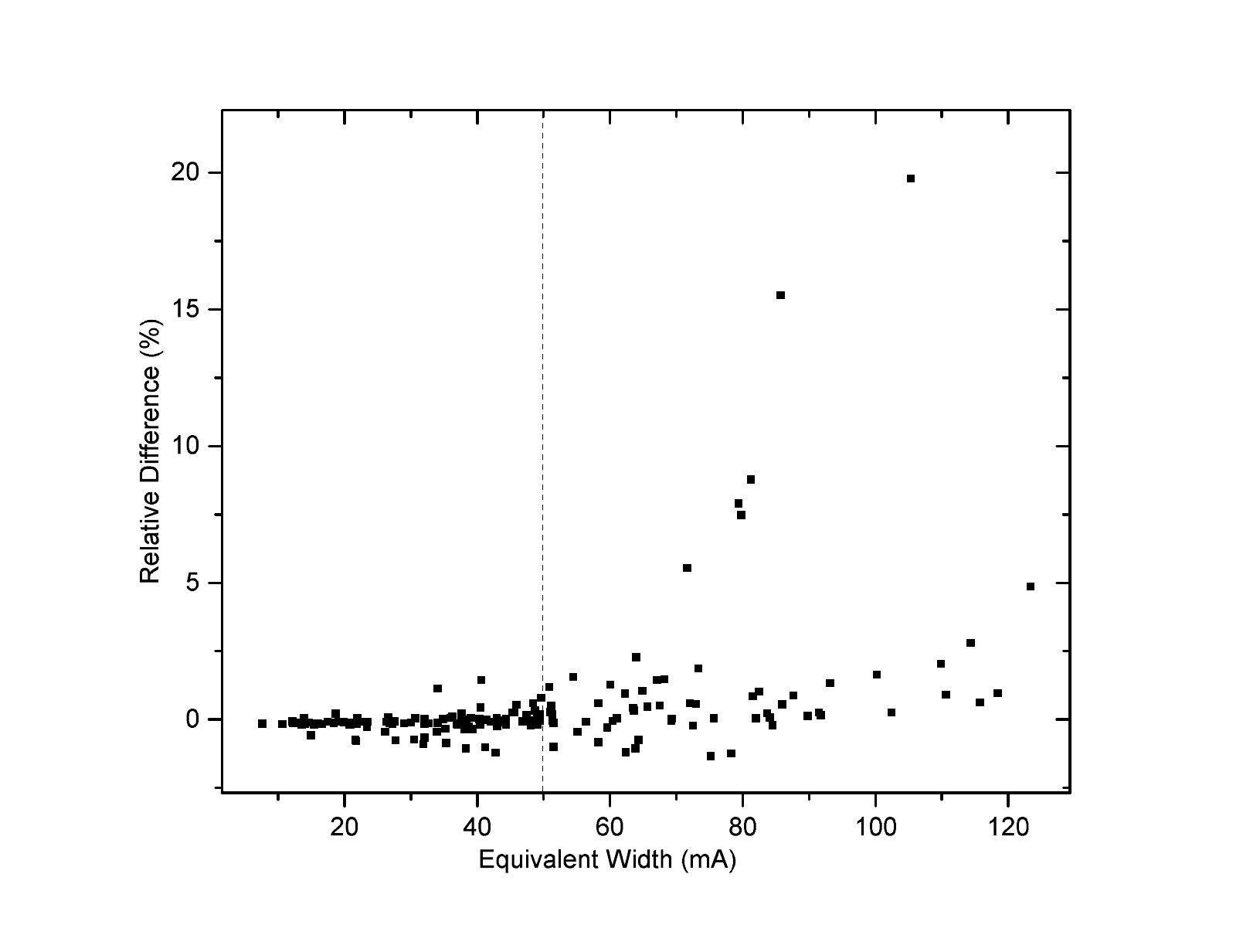}}
\caption{Relative Difference (RD) in percentage for the highest-activity model 17018 relative to the lowest-activity model 1701, versus the EW for the lines that match with those in \protect\cite{spi20}. The dashed line indicate an EW=50 m\AA. }
\label{fig:spina}
\end{figure}

\cite{spi20} observed that the EWs of lines generally increase with chromospheric activity, particularly for lines with higher EWs, with noticeable changes for those with a median EW greater than 50 m\AA. 
Figure \ref{fig:spina} explores this result, although using a different activity indicator. We plotted the RD for model 17018 against the EW of matching lines reported in their Table 2. We observed that at around 50 m\AA, indicated by a vertical line in the plot, the spread in RD begins to increase in absolute value, displaying a comparable trend to that found by \cite{spi20}, in the sense that the majority of the lines in this range show variation. Additionally, a subset of six lines exhibit an even sharper increase, marking them as highly sensitive to activity. These lines include Fe I (4603.2899 \AA, 4995.5223 \AA) and Fe II (4509.553 \AA, 4521.492 \AA, 5199.024 \AA, and 5236.082 \AA). It could be possible that our most active model represents a star that is more active than the most active star in the sample of \cite{spi20}. 

\begin{figure}[h]
\centering
\resizebox{\hsize}{!}{\includegraphics[trim=2cm 1cm 2cm 2cm, clip, width=1.0\textwidth]{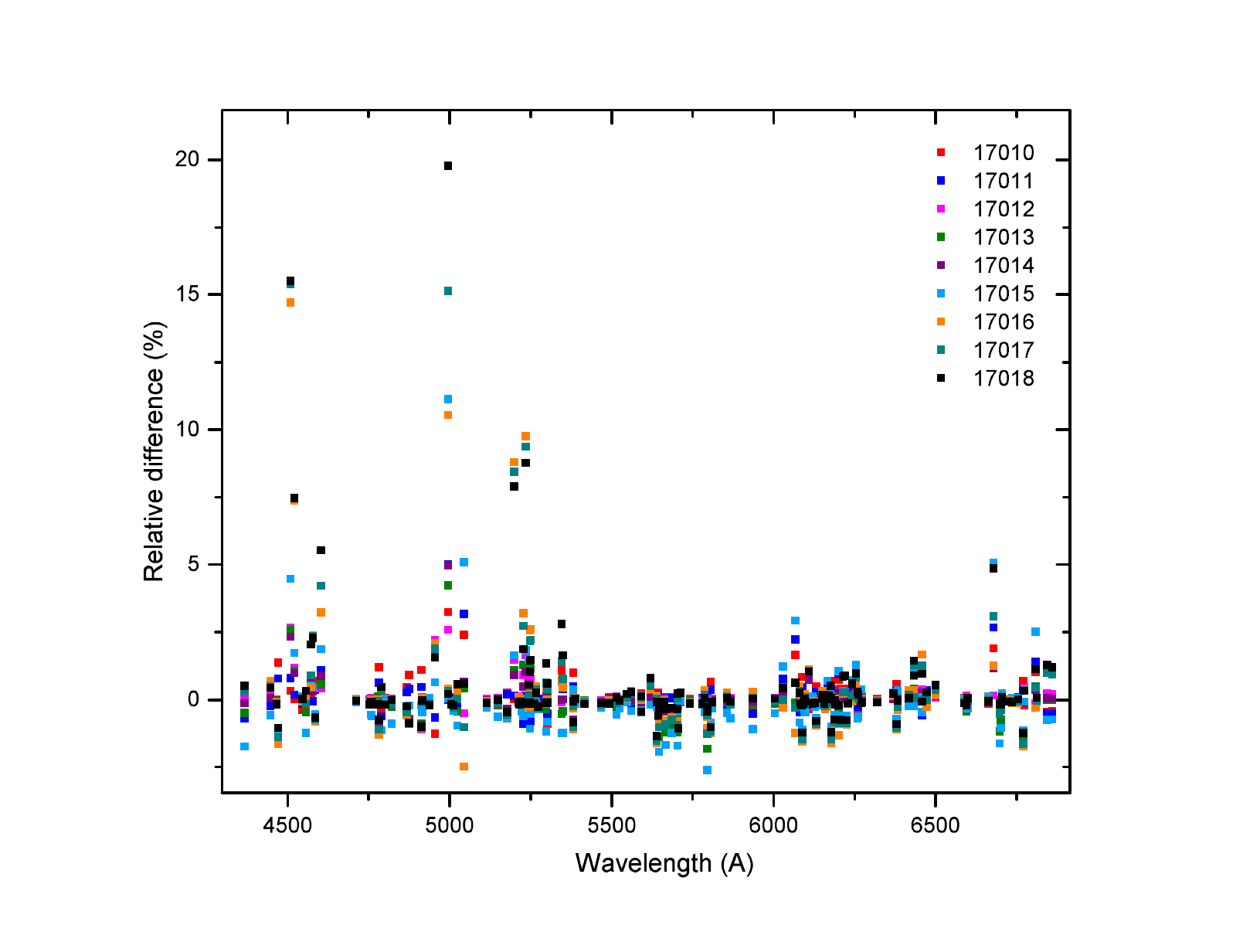}}
\caption{ RD of the matching lines with the sample of \protect\cite{spi20} produced by each model relative to the lowest-activity model 1701, as a function of wavelength (vacuum). }
\label{fig:spina2}
\end{figure}

It is important to note that the lines in their sample are the ones employed in \cite{mel14}, which was assembled specifically for the analysis of solar twin stars. They therefore preferentially selected unsaturated lines with minimal blending in the quiet sun spectrum. The rationale behind this selection highlights the importance of analyzing their behavior under increasing stellar activity. Figure \ref{fig:spina2} displays the RD values produced by each model for the matching lines in their sample as a function of wavelength. Only few lines exhibit RD values higher than 5\% for the four most active models, while the rest show smaller changes. This ensures that the less active subset of these lines can be confidently used for measuring stellar abundances or other determinations, depending on the spectral resolution.

\begin{figure*}[t]
    \gridline{
        {(a)}{\includegraphics[trim=2cm 1.4cm 3cm 2cm, clip, width=0.49\textwidth]{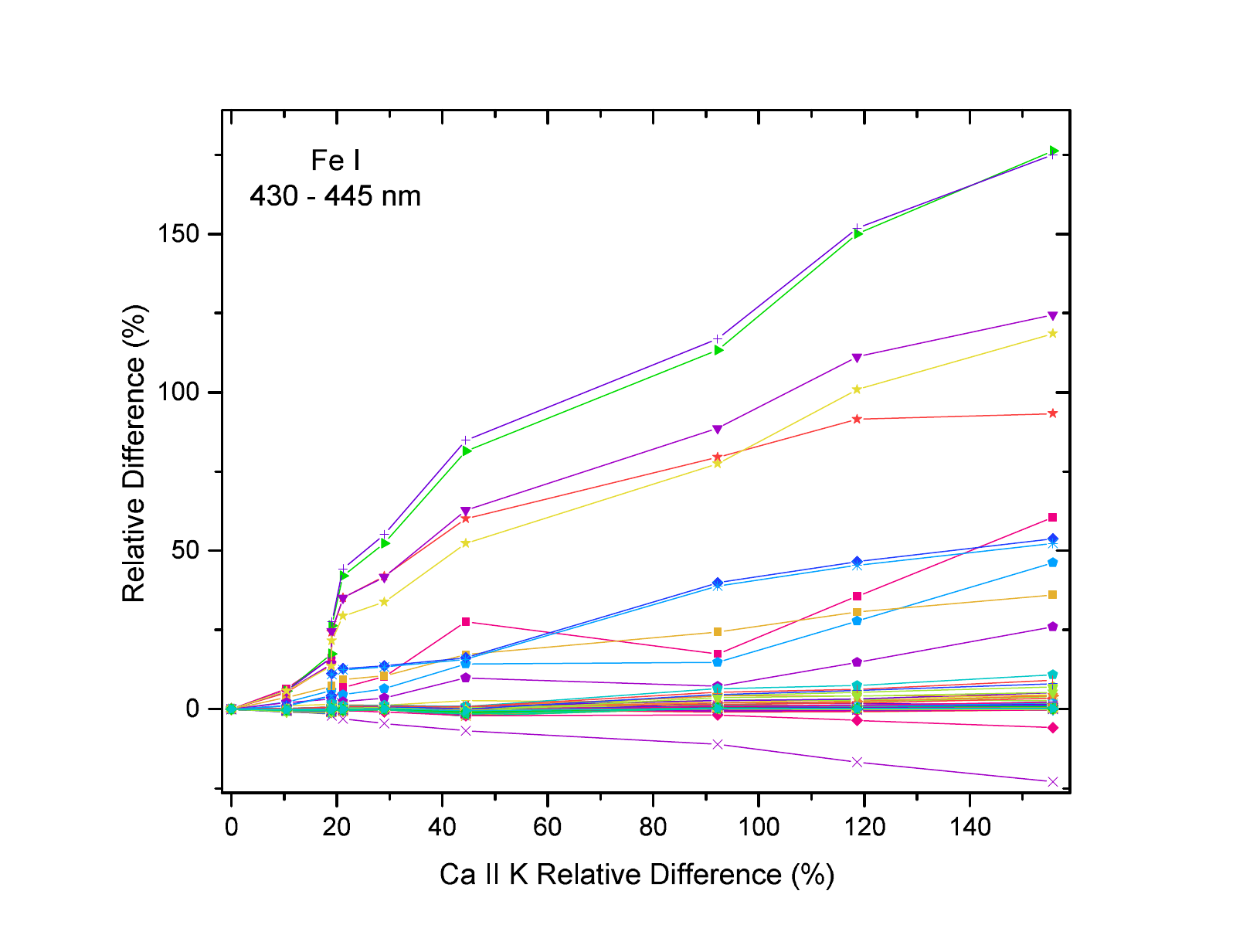}}
      {(b)} {\includegraphics[trim=3cm 1.5cm 3cm 1cm, clip, width=0.47\textwidth]{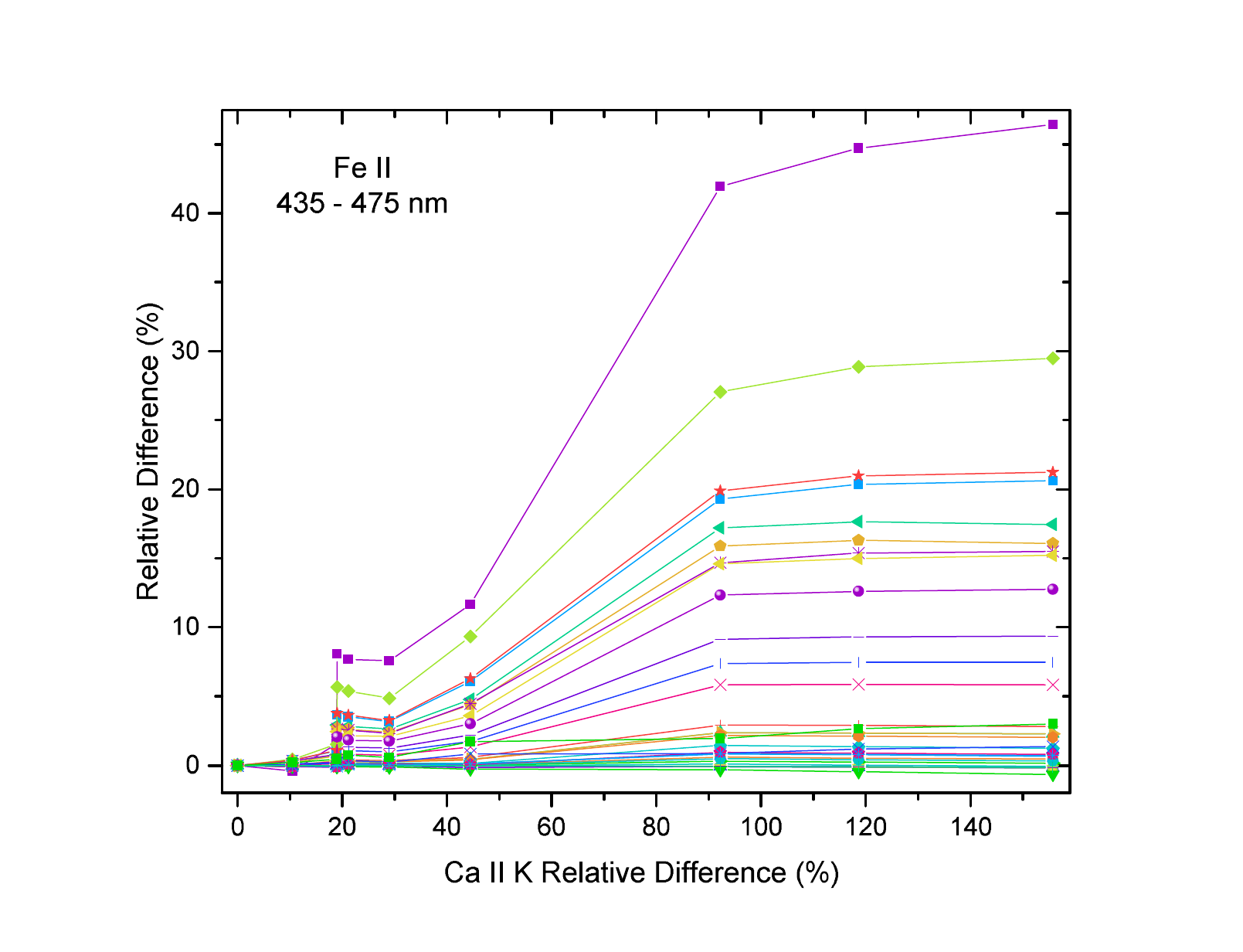}}
    }
    \gridline{
        {(c)}{\includegraphics[trim=2cm 1.5cm 3cm 2cm, clip, width=0.49\textwidth]{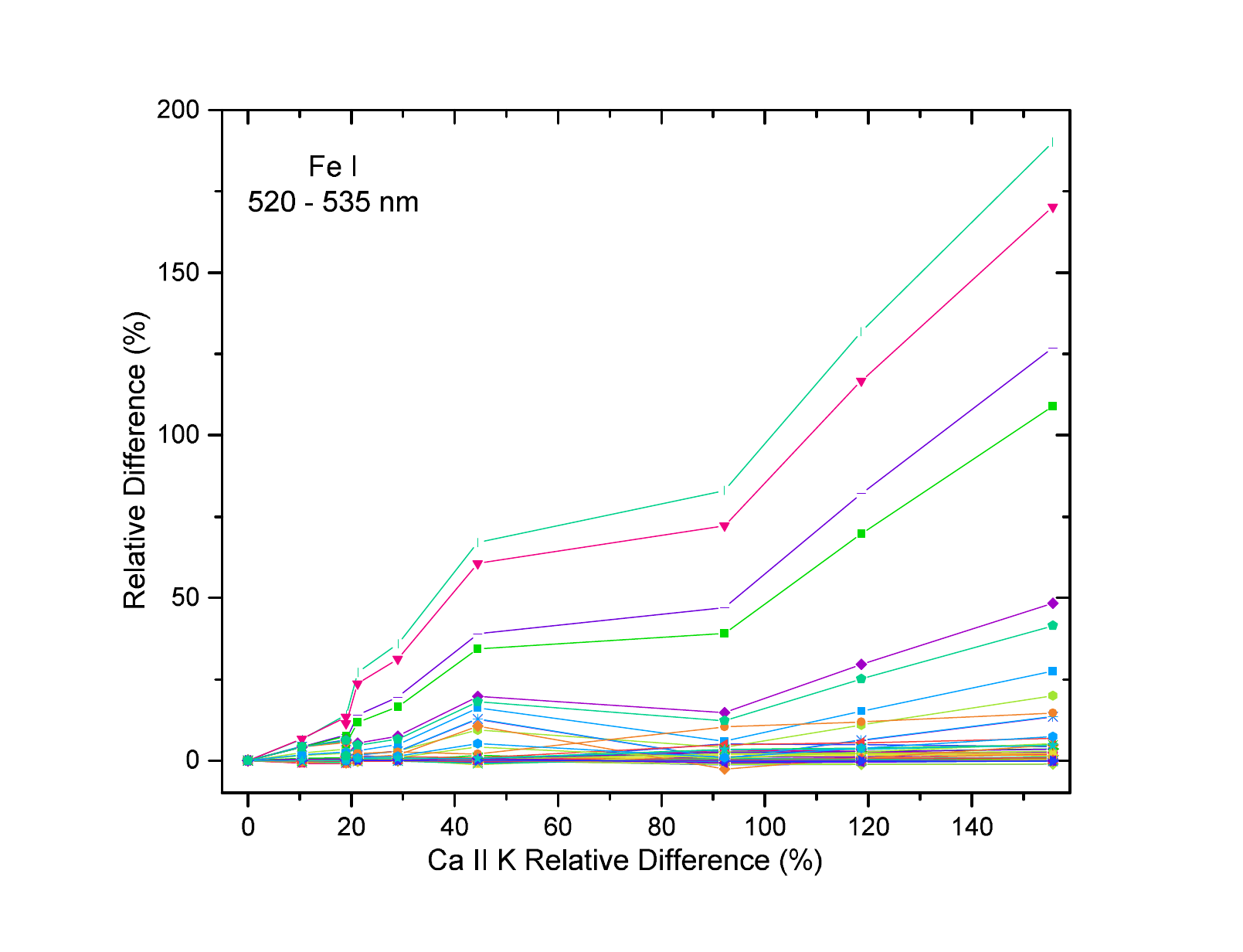}}
        {(d)}{\includegraphics[trim=3cm 1.5cm 3cm 2cm, clip, width=0.47\textwidth]{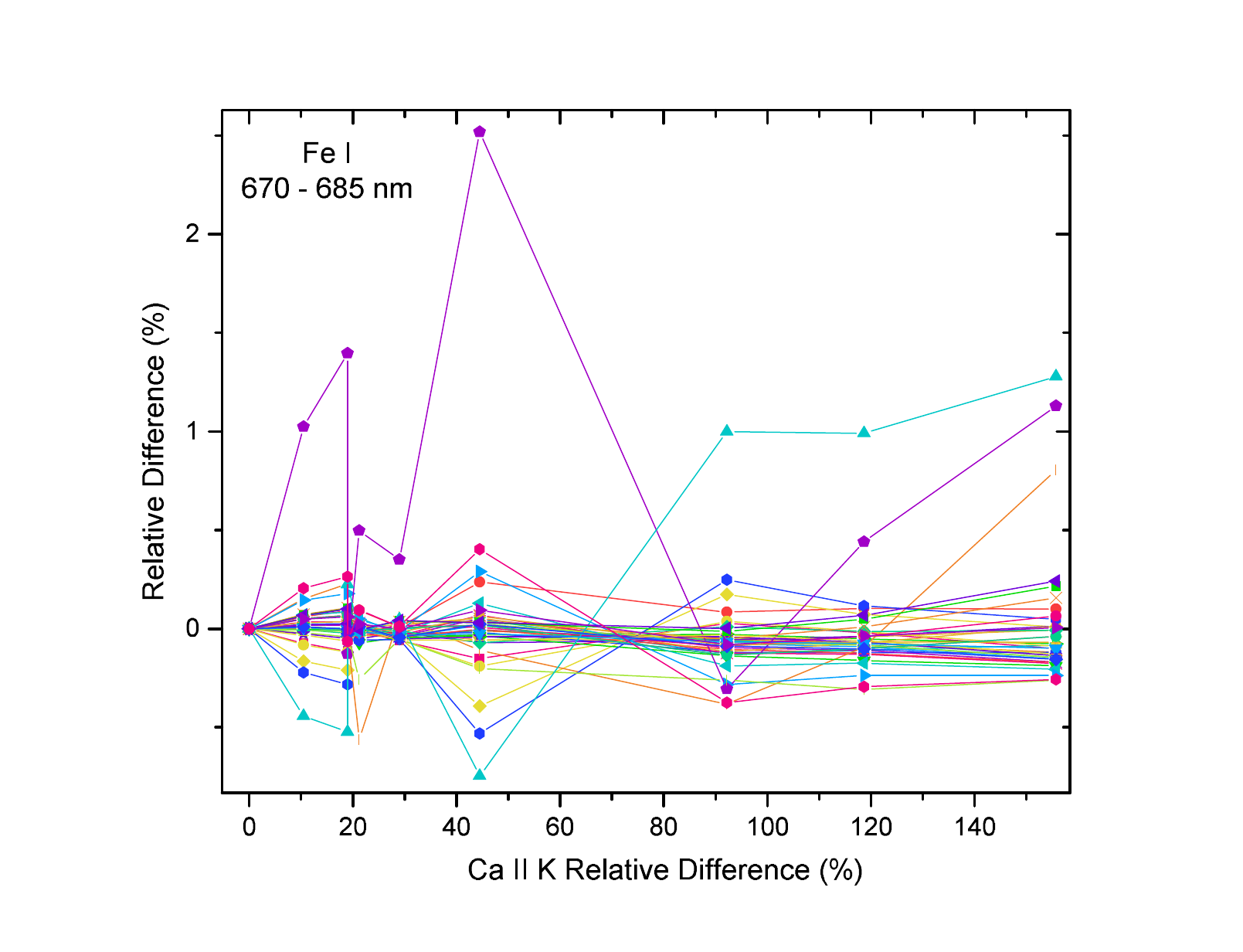}}
    }
    \caption{RD of our calculated lines with each model related to the lowest-activity model (1701) in the same spectral regions studied by \protect\cite{dra24}, as a function of the percentage RD of the integrated Ca II K line flux for each model related to the same model 1701 (shown in the fifth column of Table \ref{tab:models}). This last RD for the Ca II K line represents a measure of the increasing stellar activity. The behavior of individual lines is displayed with a distinct colored curve for easy identification.}
    \label{fig:dravins}
\end{figure*}

Using observations of the Sun-as-a-star obtained by HARPS-N during 2016, 2017, and 2018 (a period of decreasing solar activity), \cite{dra24} investigated line-strength variability associated with chromospheric activity. Figure \ref{fig:dravins} displays the same spectral ranges for Fe I and Fe II lines analyzed in their study (specifically, their Figure 6) alongside the RD of the full-disc integrated Ca II K line (see Table \ref{tab:models}). They found that the absorption strength of Fe II lines (520-535 nm) was the most affected by chromospheric activity, followed by Fe I lines in the ranges 430-445 nm and 520-530 nm, which showed similar variations. 

Our results reveal comparable trends. However, in our case, the Fe I lines in the 520-535 nm range (Figure \ref{fig:dravins} (c)) are the most positively impacted by chromospheric activity. This is closely followed by the Fe I lines in the range 430-445 nm (Figure \ref{fig:dravins} (a)), with the exception of one line identified as Fe I 4327.157 \AA~which shows an opposite behavior (indicated by the violet line with X markers at the bottom). The Fe II lines exhibit a smaller degree of change (Figure \ref{fig:dravins} (b)). In each of these cases, there is an abrupt shift in the RD when the integrated Ca II K flux reaches 20 \% at the transition between models 17012 and 17013 (see Table \ref{tab:models}). The Fe I lines in the 670-685 nm range, consistent with \cite{dra24}, show no trend with activity.

It is important to highlight two points. First, the solar chromospheric activity in their study varied less than 5 \% in the Ca II index over the period analyzed, which is significantly lower than the change observed even in our first two models. Second, we were unable to match the Fe I and Fe II lines in their Table B.1 with our line list with the same precision as the line lists of \cite{wise18} or \cite{spi20}. Given the extremely high number of Fe lines within these spectral ranges, it is possible that the lines we plotted in Figure \ref{fig:dravins} do not correspond exactly to those in their Figure 6, which could explain the observed differences.

 \cite{dra24} also studied lines of other species, such as Mg I, Na I D$_1$ and D$_2$ doublet, and the first three H Balmer lines. Several of these lines have been examined in previous works as proxies of chromospheric activity, especially in dM stars. While we replicated some of their findings in these lines, such as the flat behavior of the intercombination Mg I 457.7 nm line, our models exceed the observed range of solar chromospheric activity, showing a stronger response to activity than that reported by \cite{dra24}. To enable a more accurate comparison, we should develop a new set of models that bridge the gap in chromospheric activity levels. On the observational side, more recent HARPS-N solar observations covering the cycle maximum could be employed to increase the observed range of chromospheric activity. 
 
 Additionally, we simulated chromospheric heating in plage regions as the primary process driving the increase in the Ca II doublet, but other magnetic structures that also contribute were not considered in this study.
 
\section{Conclusions} \label{sec:conclu}
We investigated the impact of chromospheric activity on line formation by constructing a set of semi-empirical atmospheric models for hypothetical G2 dwarf stars with increasing levels of chromospheric activity, utilizing the SSRPM library of codes.

By comparing the synthetic stellar spectra of each model to the baseline model representing the Quiet Sun, we identified the most affected spectral ranges, which are approximately 3300–4400 \AA~ and 5250–5500 \AA.

Through the calculation of the contribution functions for the lines, we found that the emergence of a secondary chromospheric contribution to line formation appears to be the primary driver of these changes.

Previous observational studies have reported changes in lines of several neutral and first-ionized species in stellar spectra. Some of these studies were used to provide observational constraints on our findings. These comparisons also allowed us to validate most of the lines selected by \cite{mel14} for characterizing solar twin stars as relatively activity-insensitive features.

Based on our calculations and analysis, we have compiled a list of transition lines and their corresponding changes due to chromospheric activity. This list could serve as a valuable resource for the stellar and exoplanet research communities, facilitating improved selection of lines for stellar parameter determination or radial velocity (RV) measurements.

However, it is important to emphasize that the stellar models presented in this work were not designed to represent real stars. Instead, they were developed to analyze changes in line formation with increasing chromospheric temperature from a theoretical perspective. Creating a more detailed model for a specific star would require reliable observations of multiple spectral lines of this star to better constrain its thermal structure.

Further observational constraints, particularly from complete solar cycle data obtained through long-duration Sun-as-a-Star observing programs such as that of HARPS-N, could help validate our findings, at least at solar activity levels. Observations of stars more active than the Sun would be necessary to reach the conditions represented in our most active stellar models.

Given the unique challenges posed by understanding the effects of chromospheric activity on the spectra of M and K dwarfs, as well as other G dwarf subtypes, we plan to extend this work to include these spectral types. Additionally, our current atomic database includes species with atomic numbers up to 28 (nickel). In the future, we intend to expand this database to include heavier elements, allowing us to investigate the behavior of their spectral lines under similar conditions.

\begin{acknowledgments}
M.C. Vieytes would like to thank the Center for Computational Astrophysics (Flatiron Institute) for the Sabbatical Visiting position, during which part of this work was conducted. The Center for Computational Astrophysics at the Flatiron Institute is supported by the Simons Foundation. 
The authors would also like to thank the referee for their comments and suggestions, which significantly improved the quality of this work.
Support for this work was provided by NASA through the NASA Hubble Fellowship grant HST-HF2-51569 awarded by the Space Telescope Science Institute, which is operated by the Association of Universities for Research in Astronomy, Incorporated, under NASA contract NAS5-26555.
\end{acknowledgments}

%

\vspace{5mm}





\bibliography{activity}{}
\bibliographystyle{aasjournal}



\end{document}